\documentclass[12pt]{article}
\usepackage{amsfonts}
\usepackage{psfig}
\usepackage{latexsym}
\newcommand{\newc}{\newcommand}
\newc{\gsim}{\lower.7ex\hbox{$\;\stackrel{\textstyle>}{\sim}\;$}}
\newc{\lsim}{\lower.7ex\hbox{$\;\stackrel{\textstyle<}{\sim}\;$}}
\newc{\gev}{\,{\rm GeV}}
\newc{\mev}{\,{\rm MeV}}
\newc{\ev}{\,{\rm eV}}
\newc{\kev}{\,{\rm keV}}
\newc{\tev}{\,{\rm TeV}}
\newc{\mz}{m_Z}
\newc{\mpl}{M_{pl}}
\renewcommand{\phi}{\varphi}
\newc\order{{\cal O}}
\newc\CO{\order}
\newc\CL{{\cal L}}
\newc{\eps}{\epsilon}
\newc{\re}{\mbox{Re}\,}
\newc{\im}{\mbox{Im}\,}
\newc{\invpb}{\,\mbox{pb}^{-1}}
\newc{\invfb}{\,\mbox{fb}^{-1}}

\def\NPB#1#2#3{Nucl. Phys. {\bf B#1} (19#2) #3}
\def\PLB#1#2#3{Phys. Lett. {\bf B#1} (19#2) #3}

\def\PRD#1#2#3{Phys. Rev. {\bf D#1} (19#2) #3}
\def\PRL#1#2#3{Phys. Rev. Lett. {\bf#1} (19#2) #3}
\def\vev#1{\left\langle #1 \right\rangle}

\begin{document}

\begin{titlepage}
\begin{flushright}
{LBL-44235 \\
UCB-PTH-99/40 \\
hep-ph/9909326\\
September 1999\\
}
\end{flushright}
\vskip 2cm
\begin{center}
{\large\bf Flavor at the TeV Scale with Extra Dimensions}
\vskip 1cm
{\normalsize
Nima Arkani-Hamed, Lawrence Hall, David Smith and Neal Weiner}\\
\vskip 0.5cm
{\it Department of Physics\\
University of California \\
Berkeley, CA~~94720, USA\\ 
and \\
Theory Group\\ 
Lawrence Berkeley National Laboratory\\
Berkeley, CA~~94720, USA\\}
\end{center}
\vskip .5cm
\begin{abstract}
Theories where the Standard Model fields reside on a 3-brane, with a low 
fundamental cut-off and extra dimensions, provide alternative solutions to
the gauge hierarchy problem. However, generating flavor at the TeV scale 
while avoiding flavor-changing difficulties appears prohibitively difficult at
first sight. We argue to the contrary that this picture {\it allows} 
us to lower flavor physics close to the TeV scale. 
Small Yukawa couplings are generated by ``shining'' badly broken 
flavor symmetries from distant branes, and flavor and CP-violating 
processes are adequately suppressed by these symmetries. 
We further show how the extra dimensions avoid 
four dimensional disasters associated with light fields charged under flavor.
We construct elegant and realistic theories of flavor based on the maximal 
$U(3)^5$ flavor symmetry which naturally generate the simultaneous 
hierarchy of
masses and mixing angles. Finally, we introduce a new framework 
for {\it predictive} theories of flavor, where our 3-brane is embedded within
highly symmetrical configurations of higher-dimensional branes.
 
\end{abstract}
\end{titlepage}
\setcounter{footnote}{0}
\setcounter{page}{1}
\setcounter{section}{0}
\setcounter{subsection}{0}
\setcounter{subsubsection}{0}


\section{Introduction} \label{sec:intro}

The extreme weakness of gravity is usually attributed to the 
fundamental mass scale 
of gravity being very much larger than that of the strong and electroweak 
interactions. The standard model provides no understanding of how this 
enormous difference in scales is stabilized against radiative 
corrections.  
Despite this gauge hierarchy problem, the need for
extraordinarily large physical mass scales has been accepted as a
central feature in theories of physics beyond the standard model. The
unification of the gauge coupling constants at $10^{16} \gev$
strengthens this view. Furthermore, the absence of flavor and CP
violating phenomena, beyond that explained by the weak interactions,
has made it all but impossible to construct theories of flavor at
accessible energies, and suggests that the fundamental mass scale for
flavor physics is very far above the TeV scale.

Over the last two decades, the usual approach to addressing the gauge hierarchy
problem has been to modify particle physics between the weak and Planck scales. 
However,  there is another possibility: gravity can be
modified at and beneath the TeV scale, as was 
realized in \cite{ADD,AADD,ADD2,ADMR}. In this scenario, the fundamental mass scale of 
gravity can be brought far beneath the conventional Planck scale, 
perhaps as low as a TeV, in the presence of sub-millimeter sized 
new spatial dimensions serving to dilute the strength of gravity at long distances. 
These dimensions have not been detected since the 
Standard Model fields are localized on a three-dimensional wall, or 
``3-brane'', in the higher dimensional space.
Such a scenario can naturally be accommodated in string theory \cite{AADD}, 
where the wall on which the SM fields live can be a D-brane.  

Remarkably, despite the profound modifications of physics both at 
sub-millimeter and TeV scales, this scenario is not excluded by any known 
lab, astrophysical or cosmological constraints \cite{ADD2}. 
This realization 
opens up the possibility that there may be a number of experimentally viable 
approaches to addressing the hierarchy problem which 
involve the basic ingredients of modifying gravity at or beneath the
TeV scale,
and localizing matter fields to branes in extra dimensions.
An interesting modification of gravity has been proposed recently \cite{RS1} 
where the gravitational metric describing the 4 usual coordinates of
spacetime depends on the location in the extra dimensions. Such
metrics result from solving Einstein's equations in the presence of
brane configurations, and lead to spatial localization of the graviton
zero mode in the extra dimensions. Various schemes for solving the
hierarchy problem have been based on this \cite{RS1, RS2, ADDK, LR}. 
All these schemes, and the original scheme with large extra
dimensions, share a common feature: we live on a 3-brane located in
the extra dimensions in which gravity propagates. From the viewpoint of
our 3-brane, the fundamental mass scale is the TeV scale, and this is
the scale at which quantum gravity gets strong.

Lowering the fundamental cutoff close to the TeV scale obliterates
the usual ultraviolet desert in energy scales. On the theoretical side, this 
seems to destroy the attractive picture of gauge coupling unification. 
More pressingly, there are in principle  dangerous effects from higher 
dimension operators now only suppressed by the TeV scale 
giving e.g. disastrously large rates for proton decay. 

However, it has been realized that the space in the extra dimensions 
replaces the old desert as the new arena in which such issues can be addressed.
For instance, the old picture of logarithmic gauge coupling unification close 
to the Planck scale may be mimicked by the logarithmic variation of classical 
fields in sets of two large dimensions \cite{ADMR2}. \footnote{Another 
approach to gauge coupling unification bases on power-law running of 
higher-dimensional gauge couplings has been discussed in \cite{DDG}.}
Furthermore, the difficulties associated with higher-dimensional operators
can also find a natural resolution using higher-dimensional locality. 
Indeed intrinsically higher-dimensional ways of suppressing proton decay
were proposed in \cite{ADD2,ADD,AS}. \footnote{In a different context, 
higher-dimensional locality has been used to ameliorate the SUSY 
flavor problem in anomaly-mediated models \cite{RSSUSY}.}
After proton decay, the most 
serious issue is that of flavor-changing neutral currents.  
Dimensional analysis suggests that the flavor scale should be
above $10^4 \tev$ from the $K_L$---$K_S$ mass difference, and greater
than $10^5 \tev$ from CP mixing in the neutral kaons. While this naive
estimate can be avoided, 
it has proved extraordinarily hard to construct theories at the TeV scale
which provide an explanation for the small flavor parameters.
It is natural to ask whether extra dimensions offer any new possibilities for 
evading these problems. 
In \cite{AD}, a higher-dimensional mechanism was proposed for 
generating the fermion mass hierarchy, and preliminary arguments were given 
to suggest that the FCNC problem could also be avoided. It is our purpose in 
this paper to extend and generalize these ideas to realistic and elegant 
theories of flavor at the TeV scale which are safe from FCNC effects. 
As we will see, {\it it is the physics of extra dimensions that allows
us to naturally bring flavor physics down to the TeV scale}.

In this paper we therefore study effective theories of flavor 
with a low fundamental mass 
scale $\Lambda$. What do we mean by ``effective theories of flavor''?
At scale $\Lambda$ there is a fundamental theory, presumably string
theory, which has some low energy effective theory. It is conceivable
that this is just the standard model; with entries in the Yukawa
matrices somehow set to the required hierarchical values, and with all
higher dimensional operators absent. We consider this unlikely, but,
since we do not know the low energy limit of string theory, we must
make some assumptions about the form of the low energy effective theory.

We assume that the effective theory beneath $\Lambda$ is based on some
symmetry group G and has an effective Lagrangian
\begin{equation}
\CL_{ef\!f} = \sum_i {c_i \over \Lambda^{p-4}} \CO_i^p
\; \; \supset \; \; \left( {\phi \over \Lambda} \right)^n f f^c H  + ...
\label{eq:leff}
\end{equation}
where $i$ runs over all
$G$ invariant operators, $\CO_i^p$, $p$ labels the dimension of the
operator and $c_i$ are unknown dimensionless couplings of order
unity. An example of an operator which leads to a small Yukawa
coupling for the fermion $f$ to the Higgs $H$ is shown.
This assumption implies that the small dimensionless parameters
of flavor physics must arise spontaneously from $\vev{\phi} /
\Lambda$, where $\phi$ is a field of the low energy theory. 
We call $\phi$ a flavon field: the effective theory must
explain why it has a vev small compared to the fundamental scale.
In an effective theory of flavor, given the symmetry
group and the field content, and in extra dimensions the brane
configuration, flavor can be understood in the low
energy theory itself.

In section 2 we discuss several difficulties encountered in building
theories of flavor at the TeV scale in 4 dimensions. Higher dimension
operators lead to flavor-changing and CP violating effects which are
hard to tame, even with a flavor symmetry. An Abelian symmetry cannot
prevent enormous $K \bar{K}$ mixing, 
and a non-Abelian symmetry results in disastrous flavor-changing
(pseudo-) Goldstone bosons.
Even the maximal flavor symmetry, $U(3)^5$, 
is not quite sufficient to protect against large electric dipole
moments for the electron and neutron. Finally, the flavon quanta are
themselves very light, and the exchange of these particles in the low
energy theory also generates disastrous four fermion operators. These
difficulties are illustrated with a $U(2)$ flavor symmetry group.

In section 2.6 we discuss the minimal $U(3)^5$ flavor structure in 4
dimensions, in which the three Yukawa matrices are each promoted to a
single flavon field. This structure has been used to argue that flavor
physics can occur at a low scale \cite{U35}. 
We show that $CP$ must also be
spontaneously broken for the fundamental scale to be under 10
\tev. However, as it stands, this minimal $U(3)^5$ structure is not an
effective theory of flavor. The flavon fields contain a hierarchy of
vevs which are simply imposed by hand and not derived from the low
energy effective theory. This could be remedied by introducing a
hierarchy of symmetry breakings at a sequence of scales beneath
$\Lambda$, as proposed by Froggatt and Nielson \cite{FN}. 
However, this would produce flavons at each of these
scales, and some would be very light indeed, and their exchange would
induce disastrous flavor and CP violating interactions. This flavon
exchange problem appears generic to effective theories of flavor with
low $\Lambda$ in 4d.

With extra dimensions, there is a new possible origin for the small
flavor parameters: symmetries which are broken strongly on some source
brane may be only weakly broken on our brane because the source
brane is distant from us \cite{AD}. This idea is explored in section 3.
It is most easily
implemented by making the flavon field, $\phi$, a bulk field, which is
coupled to a source on the distant source brane so that it has a small
vev on our brane. The obstacles to constructing effective theories of 
flavor encountered in 4d are immediately removed: there is
now an origin for the small parameters in the flavon vev --  
we have a real theory for the small parameters -- and yet this
is done without introducing light flavons, solving the flavon exchange
problem. Furthermore, with order unity breaking of the discrete flavor
group on the distant branes, the pseudo-Goldstone masses can be raised
to the TeV scale.
\footnote{An alternative way to make the pseudo-Goldstones massive is
to gauge the flavor symmetry in the bulk \cite{BD}. 
Although there are
horizontal gauge bosons, they are less dangerous than usual since they
propagate in the bulk.}

The origin of this success is to understand flavor from a
hierarchy of distances in the extra dimensions, and not from a
hierarchy of mass scales in our 4d world. 
In section 3 we also discuss another phenomenon which is generic in
theories of flavor in extra dimensions when the bulk flavon field
possesses non-linear interactions. This means that the flavor breaking
felt on our brane is sensitive to the value of the flavon field in the
bulk, not just to its value on our brane. The
``sniffing'' of flavor breaking in the bulk can lead to interesting 
phenomena. 
For example, in section 6, we study a $U(2)^5$ theory,
which incorporates features of the $U(2)$ 4d theory, and in which
sniffing plays a crucial role in symmetry breaking.

In section 4 we construct a complete realistic theory of flavor in
extra dimensions with $\Lambda$ in the region of 5 -- 10 TeV. The
flavor group is maximal, $U(3)^5 \times CP$, and the minimal set of
flavons propagate in the bulk, taking classical values which result
from shining from just three source branes. Each source brane breaks a
discrete subgroup of each $U(3)$ using only triplet vevs, and the
three source branes may be identical to each other.

In section 5 we consider a particularly simple brane configuration for
realizing this $U(3)^5$ theory: our 3-brane and the three source 
3-branes are located on a 4-brane, so that shining occurs in 1
dimension. This makes the calculation of the Yukawa matrix, and the
additional flavor changing effects from the bulk, remarkably simple.

In section 6 we study the possibility of a smaller non-Abelian flavor
symmetry, and introduce a variety of bulk flavons in a way motivated
by the observed quark spectrum. This theory illustrates some of the
possibilities opened up for flavor physics in extra
dimensions. For example, a new mechanism for suppressing the neutron
electric dipole moment is proposed.

Predictive theories of fermion masses can result if the source branes
have a symmetrical geometrical configuration, as would be expected in
a dynamical theory of brane stabilization. In section 7 we study
theories in which the 9 quark masses and mixing angles are given quite
successfully in terms of just 5 free parameters. These theories are
inherently extra-dimensional, with the precise predictions reflecting
the geometry of the brane configuration, and the location of our 
three-brane.
\section{Challenges to a low flavor scale}
In this section we summarize the major challenges to lowering the scale of 
flavor physics close to the TeV scale in theories with four spacetime
dimensions. The difficulties are mostly well-known (see e.g. \cite{ACHM}), 
but it is useful to have them collected in one place. 

\subsection{Dimensional analysis}

The most serious obstacle to lowering the flavor scale $\Lambda$ 
comes from flavor-changing neutral currents, most severely from the kaon system.
With the coefficients $c_i$ in eqn.(\ref{eq:leff}) taken to be of unit magnitude
with large phases, the bounds on $\Lambda$ coming from the operators 
contributing to $\Delta F=2$ processes ($\epsilon_K$ and $\Delta m_K,\Delta  m_D, \Delta m_B$) 
are presented in Table \ref{table:t}.
\begin{table}
\begin{center}
\begin{tabular}{||c|c|c||} \hline \hline
Process     & $O$ & Bound on $\Lambda \;\; (\tev)$ \\ \hline
$\epsilon_K $ & $(d s^c)(\bar{s} \bar{d^c})$ & $10^5$ \\ \cline{2-3}
& $(\bar{s} \bar{\sigma}^{\mu} d)^2$ & $10^4$ \\ \hline
$\Delta m_K$ & $(d s^c) (\bar{s} \bar{d^c})$ & $10^4$ \\ \cline{2-3}
&$ (\bar{s} \bar{\sigma}^{\mu} d)^2$ &$ 10^3$ \\ \hline 
$\Delta m_D$ & Analogous to above & $10^3$ \\ \cline{2-3}
& ''& $5 \times 10^2$ \\ \hline
$\Delta m_B$ & Analogous to above &$ 5  \times 10^2$ \\ \cline{2-3}
& '' & $5 \times 10^2$ \\ \hline \hline
\end{tabular}
\end{center}
\caption{Bounds on $\Lambda$ from $\Delta F=2$ processes.}
\label{table:t}
\end{table}
The bounds on $\Lambda$ from the left-right operators for $\Delta m_K,\Delta m_D$ 
are enhanced by a factor of $\sim 3$ due to the QCD enhancement in running from
$\Lambda$ down to the hadronic scale. 
There are also bounds on $\Lambda$, far above the 
TeV scale, coming from $\Delta F=1$ processes such as $\mu \to e \gamma$
and $K_L \to \mu e$. 

\subsection{Implications for model-building}
While this may suggest that the scale of flavor 
should be above $\sim 10^{4} - 10^5$ TeV, it is also possible that whatever
is responsible for suppressing the Yukawa couplings of the light  
generations also adequately suppresses the flavor-changing 
operators.
For instance, if a weakly broken flavor symmetry $G_F$ is responsible for the 
fermion mass hierarchy, the same $G_F$ could suppress the dangerous
operators. 

It is easy to see that even this idea fails for generic flavor symmetries.
The reason is that the most dangerous effects arise not directly from operators
that {\it violate} $G_F$, but rather from $G_F$ invariant operators which 
violate flavor when rotated into the mass basis. Suppose for instance 
that $G_F$ is Abelian with different charges for the first and second 
generations. Then, the flavor symmetry allows the higher dimension operators
\begin{equation}
\frac{a}{\Lambda^2} (Q_1 D^c_1)(\bar{Q}_1 \bar{D^c}_1) + \frac{b}{\Lambda^2} 
(Q_2 D^c_2)(\bar{Q}_2 \bar{D^c}_2)
\end{equation}
with $a,b \sim O(1)$ 
whereas operators of the form $(Q_2 D^c_1)(\bar{Q}_1 \bar{D^c}_2)$ will have
suppressed coefficients. Nevertheless, when we rotate the fields to go to 
the mass eigenstate basis, we generate an operator
\begin{equation}
\sim \frac{(a - b) \theta_c^2}{\Lambda^2} 
(Q_2 D^c_1)(\bar{Q}_1 \bar{D^c}_2)
\end{equation}
where we have assumed that the Cabbibo angle dominantly comes from the down
sector. Note that unless $a=b$ to high accuracy, this still forces 
$\Lambda > 10^3 - 10^4 \tev$. Having the Cabbibo angle come dominantly from 
the up sector helps, but still requires $\Lambda >10^2 \tev$. The only way
out is for $a=b$, however an Abelian flavor symmetry is not enough to enforce
equality. As claimed, we see that the central challenge is to ensure that 
$G_F$ {\it invariant} operators remain harmless when rotated to the mass
eigenstate basis, and this requires $G_F$ to be non-Abelian. 
If we ignored this issue, in other words if we assume that for some reason
these invariant higher-dimensions operators are absent or have their 
coefficients magically tuned to equality, then even Abelian symmetries are
enough to adequately suppress FCNC effects from ``directly'' flavor-violating
operators \cite{AHR}. 
For instance, for any Abelian flavor symmetry we expect to 
have the operator
\begin{equation}
\frac{(\lambda_s \theta_c)^2}{\Lambda^2}  (Q_2 D^c_1)(\bar{Q}_1 \bar{D^c}_2),
\end{equation}
and even assuming maximal phase this requires $\Lambda > 7 \tev$ if 
we take $m_s$ at the lower end of its range, $\sim 90 \mev$, as is currently
favored from the lattice \cite{Lattice}. Notice, however, that in a two-Higgs doublet theory
this bound turns into $\Lambda > 7 \tan \beta \tev$, so we can not tolerate 
large $\tan \beta$. 

In the SM, such higher dimension operators are generated by 
integrating out $W's$ at the weak scale, but 
enormous FCNC's are not generated.  This is 
because the SM gauge interactions respect the $U(3)^5$ flavor 
symmetry acting separately on each of the $(Q,U,D,L,E)$ fields, 
explicitly broken only by the
Yukawa matrices. In the $U(3)^5$ symmetric limit, all the operators are 
generated automatically with equal coefficients; this maximal flavor symmetry
is strong enough to ensure that {\it flavor symmetric} operators are harmless
when rotated to the mass basis. It is then natural to explore the possibility
that the true flavor symmetry is the maximal one $G_F = U(3)^5$, and we will
consider this possibility both in the context of four dimensions and in extra 
dimensions. We find that while it is difficult to believe in a real theory
based on $U(3)^5$ in 4D, it is easy to construct elegant theories based on
$U(3)^5$ in extra dimensions. 
 
However, there is a strong constraint, even on theories based on a $U(3)^5$ 
flavor symmetry, coming from electric dipole moments of the electron and 
neutron. Any flavor symmetry would allow an operator of the form e.g. 
\begin{equation}
\frac{e^{i \phi} \lambda_d}{\Lambda^2} (Q_1 H) \sigma^{\mu \nu} D^c_1,
\end{equation}
and if the phase $\phi$ is $O(1)$, this requires $\Lambda > 40 \tev$ from the
neutron edm \cite{BDN}. 
A similar operator in the lepton sector gives
$\Lambda > 100 \tev$ from the electron edm. While it may be 
possible to lower $\Lambda$ below the $10^4 -10^5 \tev$ barrier by imposing 
powerful enough flavor symmetries, we cannot lower it past $40 \tev$ 
without making further assumptions about $CP$ violation. We must assume that
$CP$ is primordially a good symmetry, and is broken by the same fields 
breaking $G_F$. This gives a hope that the phases in the mass and edm operators
are the same and therefore in the mass eigenstate basis there is no phase in 
the edm operator. 

\subsection{Flavor-changing goldstone bosons}

We have argued that the flavor group, $G_F$, cannot be Abelian:
controlling flavor changing effects from higher dimension operators
points to a large non-Abelian flavor symmetry group.
If $G_F$ is continuous the spontaneous breaking produces familons --
flavor-changing Goldstone bosons -- leading to the very stringent
bound $\Lambda>10^{12}$ GeV.  
Gauging the flavor symmetry allows the familons to be eaten, 
but the weakness of the breaking then tells us that there will be
horizontal gauge bosons with masses much smaller than the fundamental
scale, whose exchange leads to flavor changing problems. 
The only option is to have $G_F$ be a large, {\em discrete}, non-Abelian symmetry.
However, even this case typically is excluded by the accidental
occurence of pseudo-Goldstone bosons.
At the renormalizable level, the potential for the flavon fields only
contains a few $G_F$ invariant interactions, and this typically gives
an accidental continuous symmetry, reintroducing Goldstone bosons.
Higher order operators that respect only the discrete
symmetry will give masses to these pseudos which are
suppressed by ratios of flavon VEVs to the
fundamental scale.  Moreover, these ratios will be raised to high powers, 
given the large size of the discrete group, and  we are thus
left with extremely light
pseudo-goldstone bosons that can be produced, for instance, in
$K$ decays. 

\subsection{The flavon exchange problem}

In the low energy effective theory, beneath the fundamental scale $\Lambda$,
the Yukawa couplings are generated by operators of the form
\begin{equation}
{\cal L} \sim \left( {\phi \over \Lambda} \right)^n f f^c H
\end{equation}
If we set all but one of the $\phi$'s and the Higgs to its vev, we have an 
effective coupling to $\phi$
\begin{equation} 
\left(\frac{\langle\phi\rangle}{\Lambda}\right)^{n-1} 
\frac{\langle H \rangle}{\Lambda} 
f f^c \phi \sim \frac{m_{ij}}{\langle \phi \rangle} f_i f^c_j \phi
\label{eq:linflavon}
\end{equation}
Tree-level flavon exchange then generically 
generates flavor-changing 4-fermion operators that are suppressed only by 
the {\it flavon} mass and not by the scale $\Lambda$. Unless the flavon potentials
are fine-tuned, we expect that the flavon masses $m_{\phi}$ are of the same 
magnitude as the vev $\langle \phi \rangle$, which must be smaller than $\Lambda$
in order to produce small Yukawa couplings. Using the same interactions which
generate the Yukawa couplings, tree-level flavon exchange can generate 
dangerous 4-fermion operators.  
Of course, if the flavon masses dominantly respect the 
flavor symmetry, the induced operators will be flavor-symmetric, and if the 
flavor symmetry is powerful enough these operators may be harmless. However,
since the flavon vevs themselves break the flavor symmetry at a scale $\sim 
\langle \phi \rangle \sim m_{\phi}$, 
we expect generically that the flavon masses will have $O(1)$ flavor breaking.
The generated 4-fermi operators
are clearly most dangerous if the light generation Yukawa couplings are 
generated by a single flavon coupling. For instance, suppose that the 
$12$ element of the down mass matrix is produced by the vev of a flavon  
$\phi_{12}$. then, tree-level flavon exhange can induce $\Delta S=2$ operators
\begin{equation}
\frac{1}{\Lambda^2 \langle \phi_{12} \rangle^2} (Q_2 D^c_1 H) 
(\bar{Q}_1 \bar{D}^c_2 H^\dagger)
\end{equation}
which forces $\Lambda$ back above $\sim 10^{3} - 10^{4}$ TeV from $\Delta m_K$ and 
$\epsilon_K$. 

\subsection{An example: $G_F = U(2)$}


There is a very simple theory, with $G_F = U(2)$ \cite{U2}, 
which gives highly successful quark mass
matrices, and alleviates the FCNC problem in 
supesymmetric theories. However, this theory has been studied for the
case of very large $\Lambda$ -- what happens when $\Lambda$ is reduced
towards the TeV scale?

We study just the two light generations, which transform 
as $U(2)$ doublets $\psi^a$, where $a = 1,2$. The  flavons are in a doublet 
$\phi_a$ and an antisymmetric tensor $A_{ab}$. 
The structure of the Yukawa 
matrices follows from 
\begin{equation}
{\cal L}_{Yuk} \sim h_1 \phi_a \phi_b \psi^a \psi^{c b} H + 
h_2 A_{ab} \psi^a \psi^{c b} H.
\end{equation}
Whatever triggers a vev for $\phi$ and $A$, we can always choose a basis so 
that $\phi \propto \pmatrix{0\cr \sqrt{\epsilon}}$ and 
$A_{ab} \propto \epsilon' \epsilon_{ab}$, yielding the interesting structure
\begin{equation}
\lambda \sim \pmatrix{0 & \epsilon' \cr  -\epsilon' & \epsilon}.
\end{equation}

While placing the first two generations in $U(2)$ doublets goes a long way
in erasing dangerous flavor-changing effects, there are higher
dimension $U(2)$ invariant operators, in particular
\begin{equation}
{\cal O}_{bad} = \frac{1}{\Lambda^2} Q_a D^c_b \bar{D^{c a}} \bar{Q^b},
\label{eq:badop}
\end{equation}
which give disastrously large contributions to $\Delta m_K$ and
$\epsilon_K$, forcing $M_F > 10^5$ TeV. This suggests that more complex 
models are needed with more than one $U(2)$ factor. Before discussing such 
possibilities, however, let us proceed by assuming that for some reason this
dangerous higher-dimensional operator is ${\it not}$ present in the theory 
\footnote{In a Froggatt-Nielsen theory, for instance, 
as long as the only coupling between SM fields and the heavy 
Froggatt-Nielsen fields involve flavons, the coefficient of such an operator
can be suppressed by many loop factors.}. We now show that, even making this 
assumption, a 4d $U(2)$ theory still requires $\Lambda > 10^3$ TeV.
  
The difficulty for the 4d theory is that there are physical states lighter
than $M_F$ charged under flavor, the flavons themselves. As an illustration, 
suppose that we generate vevs for $\phi$ and $A$ via independent mexican-hat 
potentials 
\begin{equation}
V(\phi,A) \sim |\phi^* \phi - p^2|^2 + |A^* A - a^2|^2.
\end{equation}
Of course, most disastrously, we get goldstone modes from the breaking of 
the global $U(2)$, and $K \to \pi + $ familon would force all the scales 
above $\sim 10^{12}$ GeV. We should really be considering a large 
discrete subgroup of $U(2)$, and there will be other terms in the potential 
that can lift the familon masses. Even if this is done, however, we
are still left with light flavons of mass $\sim p, a$. The
tree-exchange 
of $A$ in particular 
generates 
\begin{equation}
|h_2|^2 \frac{H^* H 
(\epsilon ^{ab} Q_a D^c_b)(\epsilon_{ij}\bar{Q}^i \bar{D}^{cj})}{\Lambda^2 a^2},
\end{equation}
which contains the dangerous $(Q_1 D_2^c) {\overline Q_2 \overline
  D^c_1}$ operator. Note that the mass of $A$ is 
not $U(2)$ violating, and so we have generated a $U(2)$ invariant operator
which is nevertheless dangerous. The coefficient of the operator
is real so there is no contribution to $\epsilon_K$, but there is 
a strong constraint from $\Delta m_K$:  to produce the small $12$
entries of the Yukawa matrices we need
\begin{equation} 
\frac{a}{\Lambda} \sim \lambda_s \theta_c,
\end{equation}
leading to $\Lambda >10^3$ TeV.

\subsection{Minimal $U(3)^5$ in 4D}
As already mentioned, the largest symmetry group of the standard model Lagrangian in the 
limit of vanishing Yukawa couplings is $U(3)_Q \times U(3)_{U^c} \times 
U(3)_{D^c} \times U(3)_L \times U(3)_{E^c}$.  Imposing $U(3)^5$ on the 
underlying theory therefore gives the strongest possible symmetry
suppression 
of flavor 
violating processes in the effective theory.  In the simplest realization 
of $U(3)^5$ (which we illustrate for the quark sector alone), the
symmetry 
is broken by the VEVs of a single $\chi_u$ and a single $\chi_d$, transforming
as ($\overline{3}$ ,$\overline{3}$) under $U(3)_Q \times U(3)_{U^c}$ and
$U(3)_Q \times U(3)_{D^c}$, respectively.  The effective Lagrangian
has the form of equation (\ref{eq:leff}), where the flavor
and gauge invariant $\CO_i^p$ are constructed 
from  $\chi_u$, $\chi_d$, and standard model fields.  Fermion masses,
for example, 
come from the operators
\begin{equation}
{1 \over \Lambda} Q \chi_u U^c {\tilde H} \hspace{.5 in} {\rm and}
\hspace{.5 in}
{1 \over \Lambda} Q \chi_d D^c H.
\label{eq:mass}
\end{equation} 

Having too low a flavor scale $\Lambda$ leads to conflict with
experiment.  Strong bounds come from flavor conserving operators such
as 
\begin{equation}
| H^{\dagger} D_\mu H|^2 \hspace{.5 in} {\rm and} \hspace{.5 in}
( H^{\dagger} D_\mu H) \overline{Q} \gamma^\mu Q,
\label{eq:ew}
\end{equation}
which give anomalous contributions to the $\rho$ parameter and to
fermion
couplings to the Z boson, and require $\Lambda >$ 6 and 7
TeV, respectively\footnote{Here and below, we set the relevant $c_i=1$ 
to obtain bounds on $\Lambda$.}.  Other dimension 6 operators that
lead to similar precision electroweak limits are
listed in \cite{barbieri}.  Atomic parity violation experiments and
direct searches at LEPII for 4-lepton
couplings place only slightly
milder bounds of $\Lambda >$ 3 TeV.  

Provided these requirements due to flavor conserving 
phenomena are met, 
operators that arise due to flavor breaking are relatively safe.  
For example, it is impossible 
to construct higher dimensional operators that induce
$K-\overline{K}$ mixing using only the VEV of $\chi_d$, because
$\chi_d$ is diagonal
in the down quark mass basis\footnote{This is not exactly true, as
equation (\ref{eq:mass}) gives only the leading order pieces in the
Yukawa interactions, and leaves out operators like $Q \chi_u
\chi_u^\dagger \chi_d D^c H$, for instance.  However, in spite of the
large top Yukawa coupling, we find that these additional contributions 
are not dangerous, and we omit them from our discussion.}.  One must instead
consider operators that involve $\chi_u$, such as
\begin{equation}
{c \over \Lambda^2} (\overline{Q} {\overline \sigma}^\mu {\chi_u^\dagger \chi_u
  \over \Lambda^2} Q)^2,
\end{equation}
which in the mass basis contains
\begin{equation}
{c \over \Lambda^2}(\overline{D} {\overline \sigma}^\mu V_{CKM}^\dagger
{\overline{\lambda}_U^2} V_{CKM} D)^2,
\end{equation}
where $\overline{\lambda}_U^2={\rm
  Diag}(\lambda_u^2,\lambda_c^2,\lambda_t^2)$.  
This gives
the $\Delta$S $=2$ piece
\begin{equation}
{c \over \Lambda^2} \lambda_t^4 (V_{td}^* V_{ts})^2 (\overline{d}
{\overline \sigma}^\mu s)^2,
\end{equation}
which leads to bounds from $\Delta m_K$ and $\epsilon_K$ of 
${\Lambda} >$ .5 and 5 TeV, provided the phase of 
$c (V_{td}^* V_{ts})^2$ is of order one.  
$\Delta$S$=1$ processes give weaker bounds.

As mentioned in section 2.2, the most stringent bound
on $\Lambda$ arises because, a priori, there is no reason to expect any relations
between the phases of the $c_i$ that appear in equation
(\ref{eq:leff}).  The
Yukawa interaction
\begin{equation}
Q {\chi_d \over \Lambda} D^c H
\end{equation}
and the electric dipole moment operator
\begin{equation}
{1 \over \Lambda^2}
F_{\mu \nu} Q {\chi_d \over \Lambda}\sigma^{\mu \nu} D^c H
\label{eq:EDM}
\end{equation}
can simultaneously be made real and diagonal.  However, since these
operators' coefficients have independent phases, we should expect
that the coefficient in front of the EDM operator will be complex
in the mass basis, and generically, we
get a contribution to the neutron EDM that is too large unless $\Lambda >$
40 TeV.  To evade this bound we must require that CP is a symmetry of
the underlying theory, broken spontaneously by $\chi$ vevs. 
Because the same flavon, $\chi_d$, gives rise to both the Yukawa
interaction and the EDM operator, spontaneous CP violation guarantees
that there is no contribution 
to the neutron EDM at leading order, and
the bound on $\Lambda$ disappears\footnote{More precisely, after
taking into account higher order contributions to both Yukawa and EDM
interactions, the bound is reduced to $\Lambda>$ 500 GeV.}.

Provided that the scale $\Lambda$ is larger than roughly 7 TeV, and
that CP is broken spontaneously,
minimal $U(3)^5$ sufficiently suppresses all 
dangerous operators that arise in a spurion analysis.  
Ideally, though, a flavor symmetry should do more than simply control higher
dimensional operators; it should also accomodate a simple
understanding of fermion mass hierarchies and mixing angles.  
Unfortunately, if we insist on a low flavor scale, addressing masses
and mixings in the context of $U(3)^5$ becomes problematic.  One might 
attempt to explain
mass hierarchies by introducing a few sets
of $\chi$'s that acquire VEVs at very different scales.
However, the $U(3)^5$ mechanism for suppressing dangerous
operators requires that only a single $\chi_d$ and a single $\chi_u$
exist.  If instead there were, say, two of each, then a combination
\begin{equation}
a_1 \chi^1_d +a_2 \chi^2_d
\end{equation}
would appear in the down quark Yukawa interaction, while a different 
combination
\begin{equation}
b_1 \chi^1_d +b_2 \chi^2_d
\end{equation}
would appear in the down quark EDM operator.  There is no generic reason 
for the second combination to be real in the basis that make makes the
first combination real and diagonal (although it is reasonable to
assume that corresponding entries of the two combinations are of the same
order of magnitude), so the EDM bound $\Lambda>$ 40
TeV returns.  Similarly, the operator
\begin{equation}
{c \over \Lambda^2}\left( {Q} (c_1 \chi^1_d +c_2 \chi^2_d)
D^c\right)\left( {Q}(f_1 \chi^1_d +f_2 \chi^2_d) 
D^c\right)^\dagger
\label{eq:eps}
\end{equation}
leads to a stringent bound, $\Lambda >$7 TeV, coming from the 
CP violating parameter
$\epsilon_K$.  
Thus we are led to work with only a single set of $\chi$'s, 
whose hierarchical VEVs, we might imagine, arise due to a sequential 
breaking of the flavor group at widely separated scales.
But by adopting this view we encounter the flavon exchange problem
problem of section 2.4: one expects the masses of the 
various flavons that compose the $\chi$'s
to be of the same order of magnitude as their VEVs, and thus the masses of the
lightest flavons to be much smaller than $\Lambda$.  Unless
$\Lambda$ is quite large, these light flavons mediate flavor changing and CP
violating processes at unacceptable levels.  

\section{Small parameters from extra dimensions}


What is the origin of the small dimensionless flavor parameters of the
standard model? All attempts at understanding these numbers have been
based on two ideas. One idea is that these parameters vanish at tree level
and are generated radiatively, and that the loop factor is small. In a
perturbative theory, with coupling parameters of order unity, the loop
factor is of order $1/16 \pi^2$. The second idea is that the small
fermion mass ratios and mixing angles arise as a ratio of mass scales
of the theory, presumably generated dynamically. Such is the situation
in Froggatt-Nielsen type theories and in extended technicolor models.

In theories with extra dimensions, however, another
attractive possibility arises. Suppose there are flavor symmetries 
that are primordially exact
on our brane, but which are strongly broken on a distant brane. If
bulk fields charged under these symmetries are
present, this symmetry breaking is
``shined'' from the distant branes \cite{AD}, and there is a
new origin for the small parameters, namely, the large volume in the 
extra dimensions. The fermion mass ratios and mixing angles are small
not because of small breaking on the distant branes, but rather 
due to the flavor breaking messenger's propagation over 
large distances across the bulk. Fundamentally, the origin is again one of a
ratio of mass scales. However, these are set by the distances in the 
brane configuration, and result in completely new physics
possibilities different from other scenarios.

Effects of this shining can be grouped into two categories:
spurion effects arising from the free classical theory, and classical
and quantum ``sniffing'' effects, arising from nonlinearities in the
Lagrangian. 

\subsection{Free, classical shining}

The basic shining effect can be understood as the classical, free propagation
of the flavon field through the bulk. From the viewpoint of physics on
our brane, the flavor breaking is at this level equivalent to classical spurion
effects. We assume that there is some flavor symmetry group $G_F$,
which acts on the matter fields $Q_i,U_j,D_k,L_m$ and $E_n$, where
$\{i,j,k,m,n\}$ specify the representations under which the fields
transform. We further assume that $G_F$ is broken at order one on 
some distant brane by a source $J$\footnote{In what follows, we have taken the
  dimensionality of $J$ to be that of a scalar field living on the
  symmetry breaking brane. Later, for simplicity, we will set $J$=1.}. If
$J$ couples to some bulk field, 
then that field can mediate the flavor symmetry breaking to our
wall. For example, suppose the Lagrangian is \footnote{Here we have
  allowed that $\chi$ transform as a 
  reducible multiplet under $G_F$ for the sake of generality. In an
  actual model there may be many $\chi$ fields, each transforming as
  an irreducible multiplet.}
\begin{equation}
\mathcal L \supset \int d^4x \> dy^m_\parallel dy^n_\perp {J_{k l}
\chi^{k l} \over M_*^{n-4 \over 2}} \delta^n(y_{\perp}-y_0) + \int d^4x \>
dy^{m+n} {\chi^{k l} \over
  M_*^{m+n+2 \over 2}} L_k E_l H \delta^{m+n}(y),  
\label{eq: basiclag}
\end{equation}
where the symmetry is broken on a $(3+m)$-brane at location $y_0$ in the extra 
dimensions, and where $M_*$ is the fundamental scale . Since 
the source brane is an extensive object, it acts as
a point source for a Yukawa potential in $n$ dimensions. This is
completely analagous to a charged plate in 3 dimensions being
described as a point source in 1 dimension. Knowing this, it is simple 
to write down what the profile of the $\chi$ field is as a function of 
$y$ (neglecting nonlinear interactions), 
\begin{equation}
\chi = J \Delta(m_\chi; y) =  {J \over M_*^{n-4 \over 2} (2 \pi)^{n \over 2}} ({m \over
  |y|})^{n-2 \over 2} K_{n-2 \over 2}(m|y|)
\label{eq:chivev}
\end{equation}
where $|y|$ is the distance from the source brane to the point in
question, $K_n$ is the modified Bessel function,
 and $n$ is the codimension of the source of $\chi$ in the space
in which $\chi$ propagates. For $ m_\chi |y| \ll 1$, this takes the
asymptotic form

\begin{eqnarray}
\chi \approx   {J M_* \over 2 \pi} \log(m_\chi |y|) 
\>&& (n=2) \\  
\approx { J\>  \Gamma ({n-2\over 2}) \over 4  \pi^{n \over 2} M_*^{n-4
    \over 2} |y|^{n-2}} && (n>2)
\label{eq:chivev1}
\end{eqnarray}
and for $m_\chi |y| \gg 1$,
\begin{equation}
 \chi \approx {J\> m_\chi^{n-3 \over 2} \over 2(2 \pi)^{n-1 \over 2}
   M_*^{n-4 \over 2}} { e^{- m_\chi |y|}  \over
  |y|^{n-1 \over 2}}
\label{eq:chivev2}
\end{equation}
 In this example, the lepton Yukawa matrix will be 
\begin{equation}
\lambda_l^{mn} = {1 \over M_*^{(n+m+2)/2}} \chi^{mn}.
\label{eq:example1}
\end{equation}
The flavor symmetry breaking parameters are then either
power or exponentially suppressed functions of the distances between
branes. If the different elements of $\chi$ are generated on different branes, 
we can, at least in principle, generate a fully general texture, and
likewise for the quarks. For example, with $G_F = U(3)^5$ the flavon
fields $\chi_{u,d,e}$ appear as single multiplets on our wall, and yet
the various entries can have values which are hierarchical.

\subsection{Classical and quantum sniffing}

Flavor breaking from extra dimensions is much more interesting 
than simply taking the values of $\chi$ and its derivatives on our
brane at $y=0$ and using them in a spurion analysis.
Non-linearities in the bulk Lagrangian can induce a wide
variety of effects which probe flavor breaking at non-zero $y$.

The simplest examples of this are classical non-linear effects. One of 
the most significant is the generation of a vev for a
bulk field without a direct source brane. Consider a situation with two
source branes, with sources $J^1_m$ and $J^2_n$ and two bulk fields
which have vevs generated on these branes, $\chi_1^m$ and
$\chi_2^n$. If there is, in addition, another bulk field $\phi^{mn}$
transforming as a product representation of the 1 and 2
representations, we naturally have a term in our Lagrangian
$\phi^*_{mn} \chi_1^m \chi_2^n$. As a consequence, $\phi$ will also
take on a vev in the bulk, and hence on our wall as well.

This is a very familiar situation, even in four dimensions. However,
in four dimensions, we typically expect a value $\phi \propto \chi_1 
\chi_2$. In extra dimensions, the vev can typically be much
larger. As shown in Fig. \ref{fig:sniff}, the fact that the source for 
$\phi$ is spread throughout the bulk means that
the dominant contribution to its shining can come from a region distant 
from our brane, defeating our four-dimensional intuition.

Formally, we want to sum the contributions to $\phi$ on our wall
from every point in space. Given the assumed coupling, if we take our
wall at $y=0$, and the sources for $\chi_1$ and $\chi_2$ at $y_1$ and
$y_2$, respectively, we can then calculate
\begin{equation}
\phi(0) = \int d^n y \Delta(m_\phi; |y|)
\chi_{1}(y) \chi_{2}(y).
\label{eq:sniff1}
\end{equation}
If the propagators are dominantly exponentials, some small region will 
dominate this integral, and we can take 
\begin{equation}
\phi(0) \approx \Delta V \Delta(m_\phi; |\overline y|)
\chi_{1}(\overline y) \chi_{2}(\overline y),
\label{eq:sniff2}
\end{equation}
where $\overline y$ is some representative point within the volume
$\Delta V$ where the integrand is appreciable. In cases where the
particles are light, $\Delta V$ can be very large. Even in cases where 
the particles are heavy, if they are even an order one factor lighter
than the fundamental scale, the volume will typically be larger than
one by a factor $1/m_\chi^n$.
 
We illustrate this with the following example:
\begin{figure}
  \centerline{ \psfig{file=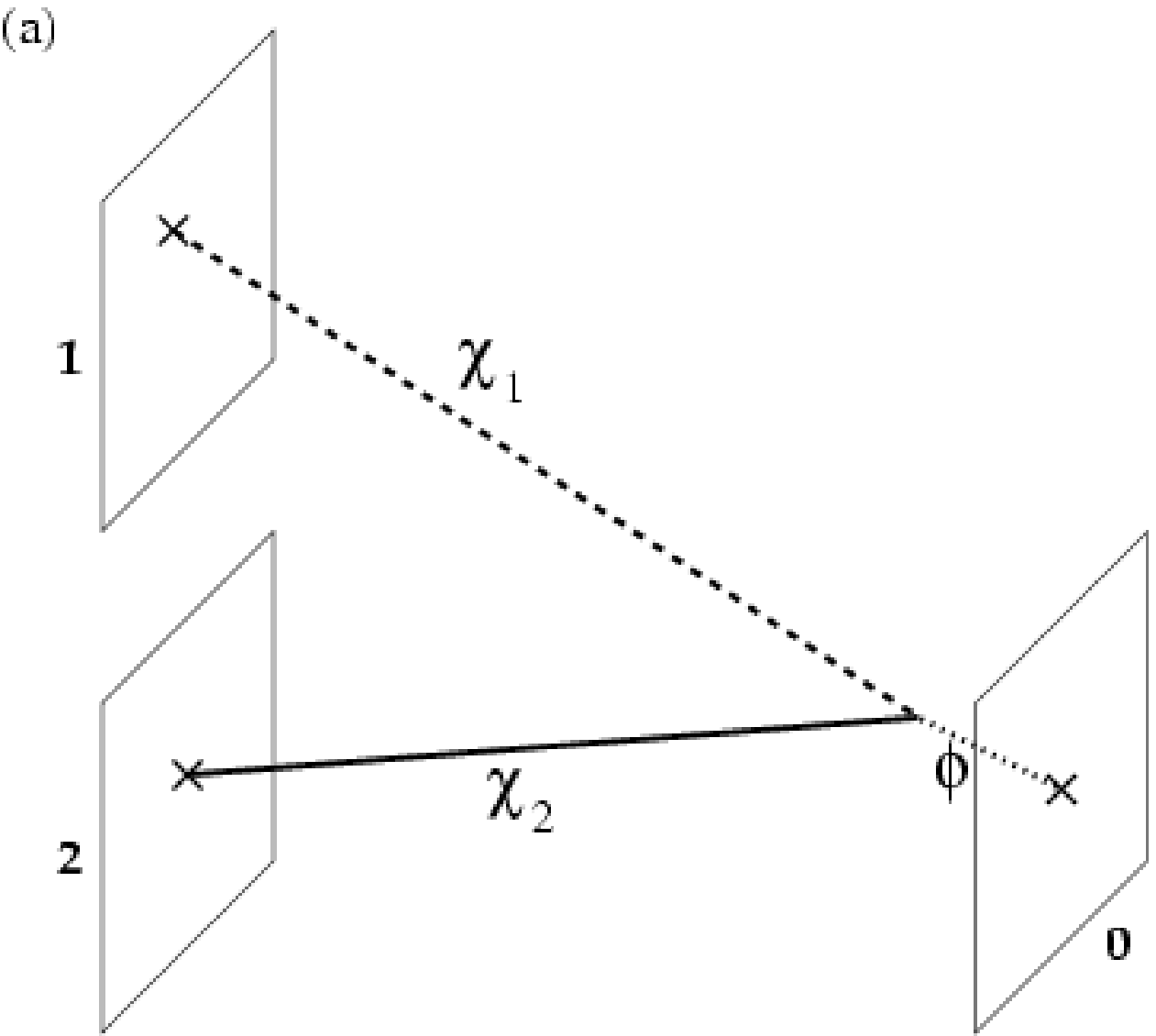,width=0.45\textwidth,angle=0} \hskip 
    0.25 in
    \psfig{file=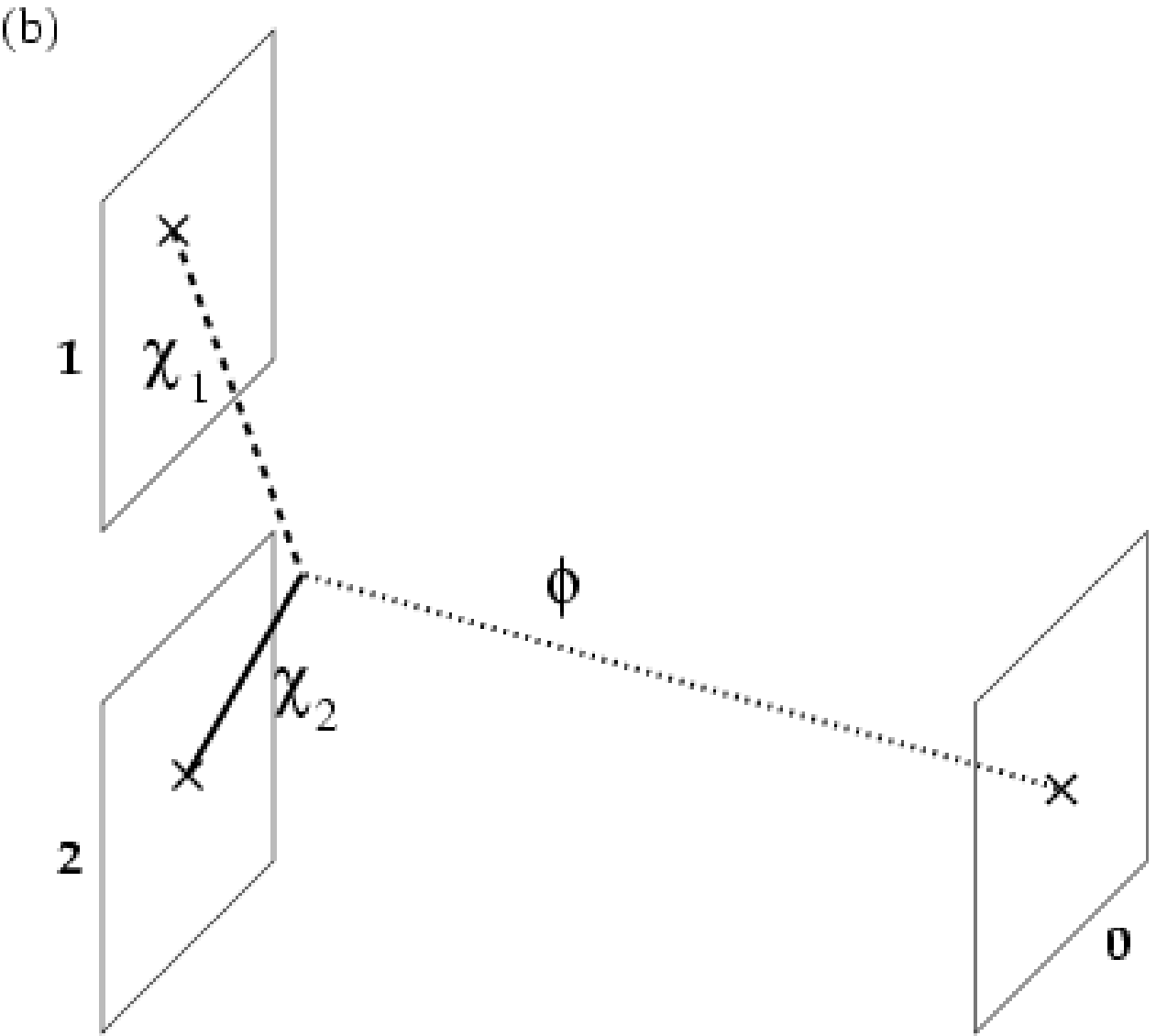,width=0.45\textwidth,angle=0} }
        \caption[sniff]{Contributions to the vev of the field
          $\phi$. Our brane is designated 0, while the source branes
          for $\chi_1$ and $\chi_2$ are 1 and 2, respectively. 
          (a) is supressed by two propagators while (b) is
          suppressed by only one. As a consequence, (b) will typically
          dominate. If the mass of $\phi$ is even an order one factor
          lighter than either of the other fields, the difference can
          be further amplified.}
        \label{fig:sniff}
\end{figure}

\begin{figure}
  \centerline{ \psfig{file=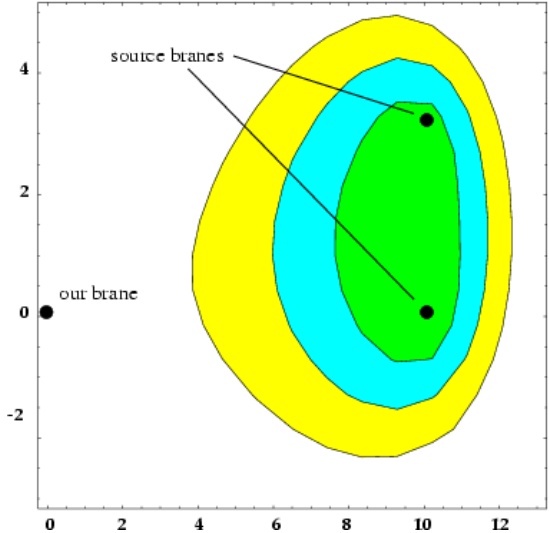,width=0.45\textwidth,angle=0} }
        \caption[sniffvol]{We plot here a $z=0$ slice of the first,
          second and third efolds of the integrand  
          of (\ref{eq:sniff1}) from its maximum for the given
          example. Notice that the region contributing to the integral 
          is both large ($\Delta V >1$) and far from our brane.}
        \label{fig:sniffvol}
\end{figure}

Consider a brane configuration with our brane localized at $(0,0,0)$ in
three extra dimensions, while one source brane is at $y_1=(10,0,0)$ and
the other at $y_2 = (10,3,0)$ in units where $M_* = 1$. Further, take
the masses to be $m_{\chi_1} \approx m_{\chi_2} \approx m_{\phi} \approx
1/3$. We can calculate the vev of $\phi$ numerically, and find on our
brane we have $\chi_1 \approx 2 \times 10^{-4}$,
$\chi_2 \approx 2 \times 10^{-4}$, and $\phi \approx 3 \times
10^{-6}$. We can understand the larger value of $\phi$ as also being 
enhanced by a volume factor $\Delta V$ of (\ref{eq:sniff2}) being
larger than one, and we show this graphically in Fig. \ref{fig:sniffvol}.

 If we further give the fields moderately different
masses, $m_{\chi_1} \approx 1/3$, $m_{\chi_2} \approx 1/2$, $m_{\phi} 
\approx 1/5$, we find $\chi_1 \approx 2 \times 10^{-4}$,
$\chi_2 \approx 5 \times 10^{-5}$, but $\phi \approx 
10^{-5}$ , much larger than the naive expectation if all other masses
are order the fundamental scale.
This is very sensitive to the brane geometry and the masses of the
bulk fields, of course, and the predictivity suffers as a
consequence. However, it illustrates one very important difference
between flavor model building in extra dimensions versus that in the
usual four.

\begin{figure}
  \centerline{ \psfig{file=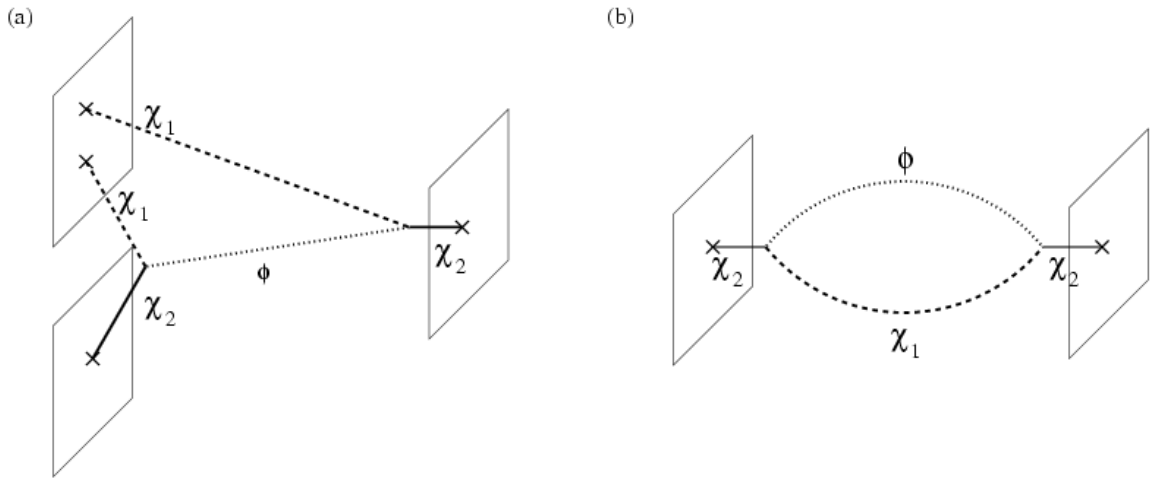,width=0.95\textwidth,angle=0}}
        \caption[sniff]{In addition to the standard direct propagation 
          of $\chi_2$, there can be other contributions. In the
          presence of an external field $\chi_2$, it can be
          advantageous to propagate as a $\phi$, as shown in (a). Even 
          absent such an external field, the propagation through a loop of
          lighter particles may be dominant compared to the 
          direct propagator (b).}
        \label{fig:nlints1}
\end{figure}

Once a bulk vev for a field exists, it can act to further regenerate
another field that may fall off more quickly. In the previous example, 
if we reach a region where $\chi_1 \phi > \chi_2$, then
this region can act as a further source for $\chi_2$, dominating for 
some regions of $y$ over the source brane. This can be understood
rather simply: in situations where the vev profile is dominantly
exponential, (i.e., when $m_{\chi_2} y \gg 1$), and if further
$m_\phi < m_{\chi_2}$, it can be advantageous to exploit
the presence of $\chi_1$, for instance, and propagate as a
$\phi$, as we illustrate in \ref{fig:nlints1}(a).  Of course, if the 
regenerated value of $\chi_2$ is sufficiently large, it can again 
regenerate $\phi$ in certain circumstances and the nonlinearities can
dominate the entire problem. It is important to be aware of this when
employing non-linear effects in model building. 

All of the effects
discussed so far arise from classical field theory in the bulk. They
could be obtained, in principle, by solving the non-linear classical
fields equations.\footnote{There is one remaining classical effect,
  namely the generation of local operators through bulk
  non-linearities, which we will discuss at the conclusion of this
  section.}
 Quantum effects in the bulk may be just as, if not 
more, important. For example, if $m_\phi + m_{\chi_1} < m_{\chi_2}$,
it may be advantageous to pay the price of a loop factor and propogate 
as a $\phi-\chi_1$ loop, which is shown in the ``Tie-Fighter''
diagram of figure \ref{fig:nlints1}(b).

\begin{figure}
  \centerline{ \psfig{file=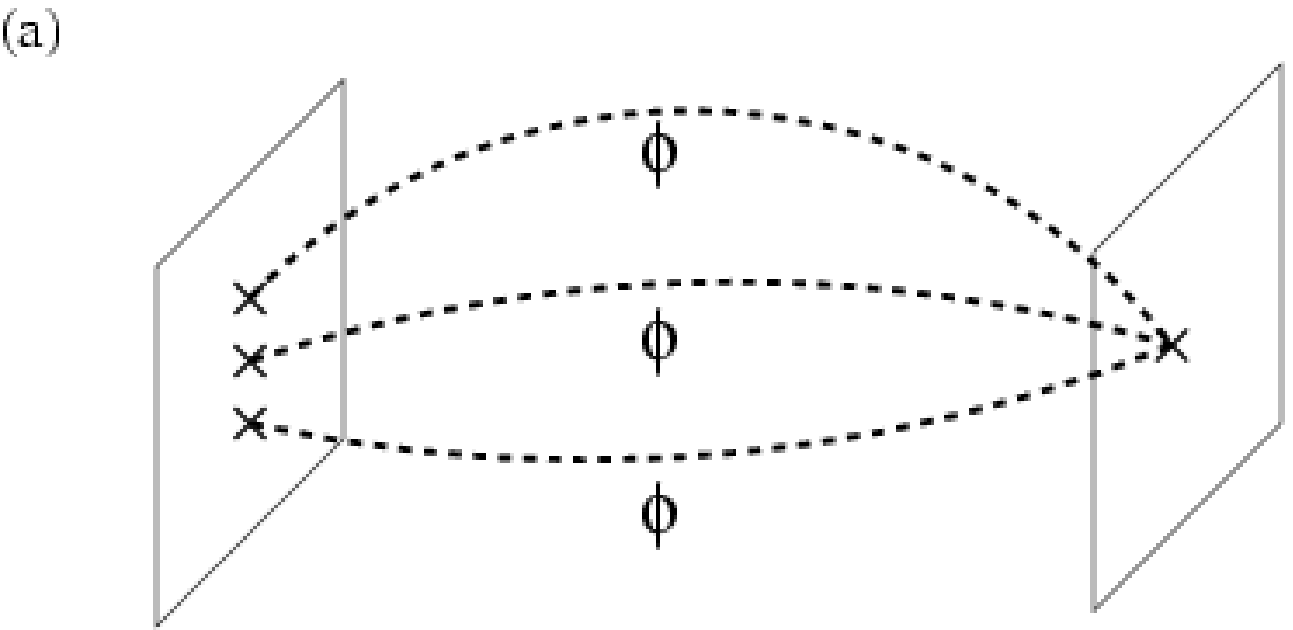,width=0.45\textwidth,angle=0} \hskip 
    0.25 in
    \psfig{file=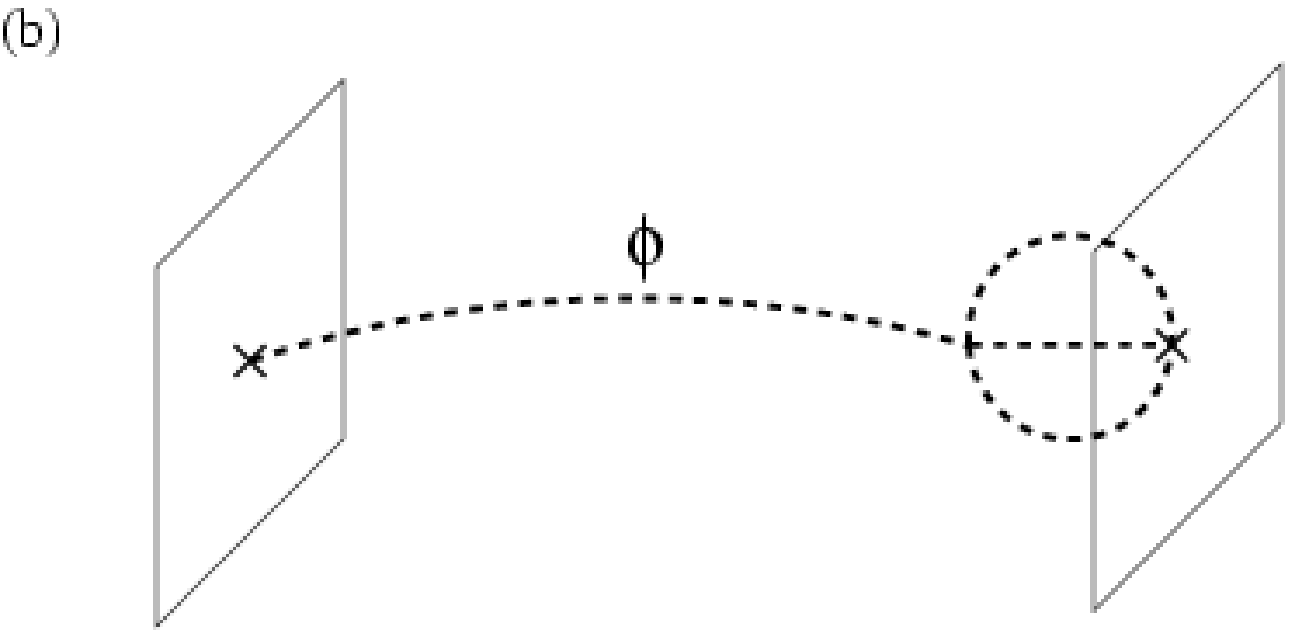,width=0.45\textwidth,angle=0} }
        \caption[sniff]{Two contributions to the value of $<\phi>^3$ on
          our brane. The contribution in (a) is the classical spurion
          contribution. The contribution in (b) is due to sniffing and 
          can often be larger than that of (a).}
        \label{fig:nlints2}
\end{figure}

As another example of non-linear bulk interactions changing the
physics on our brane at $y=0$, consider a bulk field $\phi$ with a
$\phi^4$ interaction and a source brane at $y_s$. Ignoring the
non-linearity, an operator involving $\phi^p$ on our wall would have
classically a coefficient of $\Delta(m_\phi; y_s)^p$. This is
illustrated in \ref{fig:nlints2}(a) for $p=3$. However, if $m_\phi
|y_s| \gg 1$, there will be a large exponential suppression of this
contribution, so that the dominant effect may instead come from 
the loop diagram of \ref{fig:nlints2}(b), which gives
a contribution to $<\!\!\phi(0)^3\!\!>$ of
\begin{equation}
L^2 \int d^n y \> \Delta(y)^3 \phi(y),
\label{eq:qsniff}
\end{equation}
where $L$ is a loop factor
\footnote{Actually, the exact expression involves an integration 
both over 4 and higher dimensional momenta. The result can be re-expressed 
(upon wick rotating to Euclidean space) as 
\begin{equation}
\int \frac{d^4 k_4}{(2 \pi)^4}\frac{d^4 \overline k_4}{(2 \pi)^4} 
\int d^n y \Delta(\sqrt{m_\phi^2 + k_4^2},y) \Delta(\sqrt{m_\phi^2 + 
\overline k_4^2},y)  \Delta(\sqrt{m_\phi^2 + (k_4+\overline k_{4})^2},y) 
\phi(y)
\end{equation} 
In our expression in the text, we are {\it overestimating} this effect by 
replacing $\Delta(\sqrt{m_\phi^2 + k_4^2},y)$ with 
$\Delta(m_{\phi},y)$. Henceforth, we will often make similar approximations 
in discussions of sniffing.}. 
This quantum effect can lead 
to a very large deviation of $<\!\!\phi(0)^3\!\!>$ from
$<\!\!\phi(0)\!\!>^3$, since it involves only one power of $e^{-
  m_\phi |y_s|}$. 

\begin{figure}
  \centerline{ \psfig{file=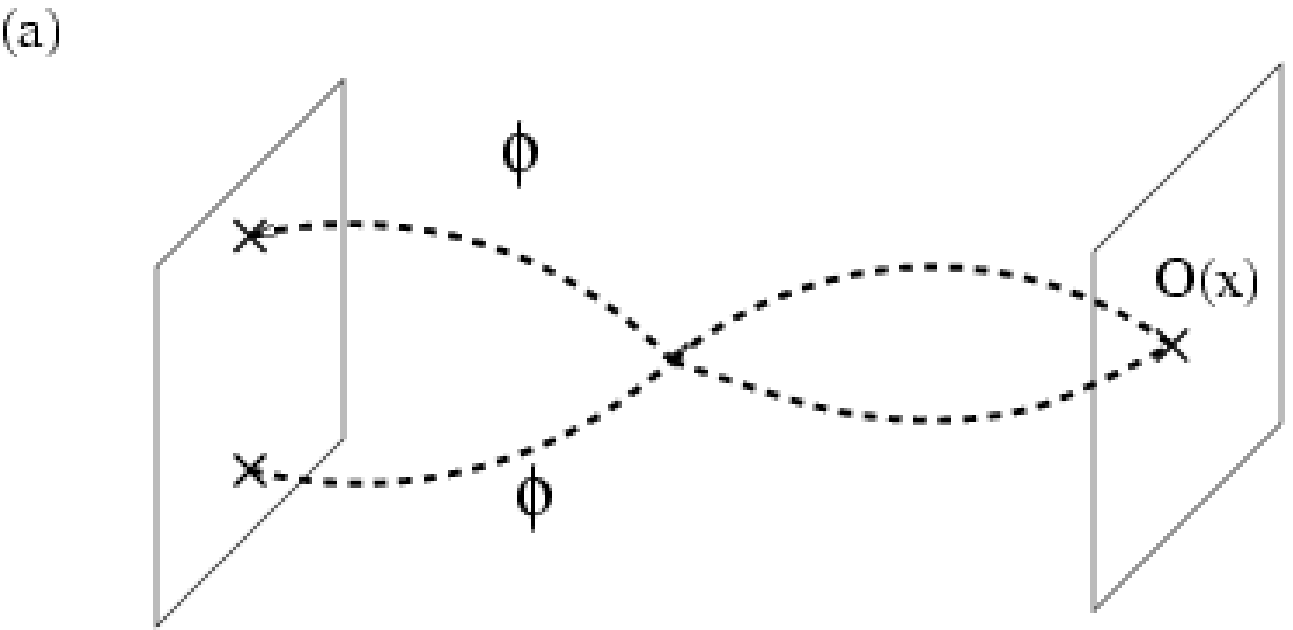,width=0.45\textwidth,angle=0} \hskip 
    0.25 in
    \psfig{file=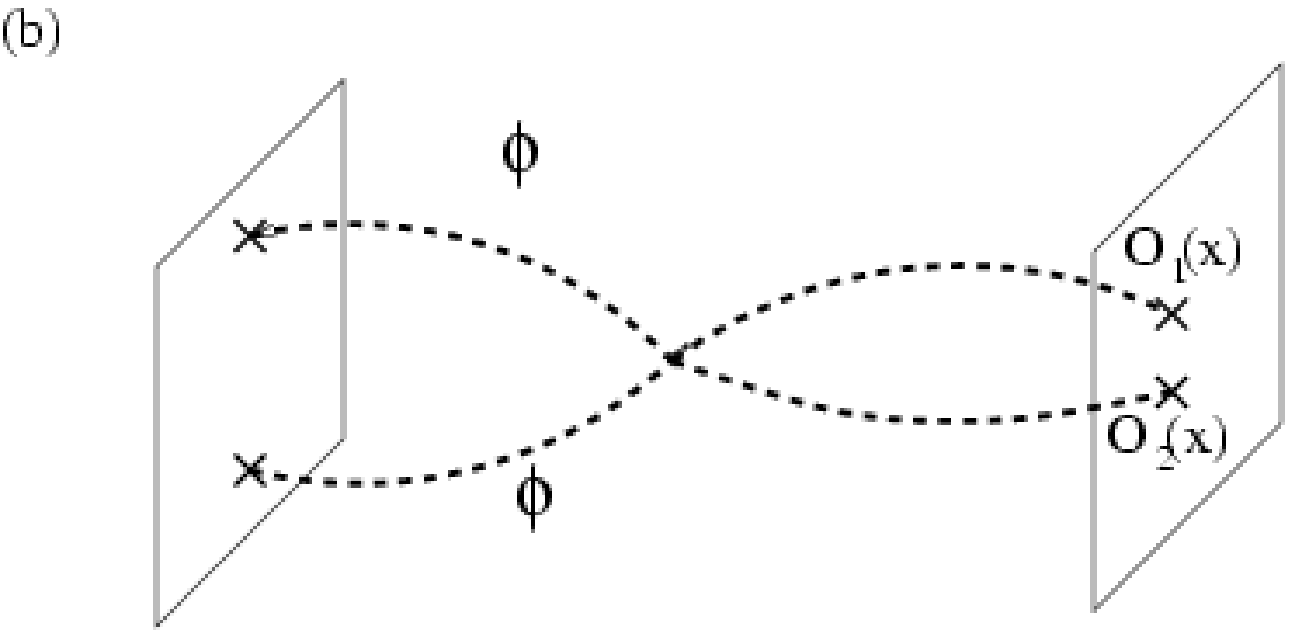,width=0.45\textwidth,angle=0} }
        \caption[sniff2]{Two examples of ``sniffed'' contributions to
          an operator on our wall. In (a) a quantum loop corrects the
          value of $<\!\!\chi^2\!\!>$ on our wall. In (b), a $\phi$
          field can interact in the bulk and generate a local
          operator.}
        \label{fig:nlints3}
\end{figure}

For the case of $p=2$ and an operator on our brane involving
$\phi^2$, in addition to the tree contribution there is the 1-loop contribution
shown in Fig. \ref{fig:nlints3}a, equal to

\begin{equation}
L \int d^n y \> \Delta(y)^2 \phi(y)^2.
\label{eq:qsniff2}
\end{equation}
In this case, two propagators traverse the whole space from $y=y_s$ to 
$y=0$ in both tree and loop diagrams, so it may appear that the loop
correction is small and unimportant. However, in theories of flavor, we will
find that the loop diagram contains flavor breaking not present at
tree level. For instance, if $\phi$ transforms as a multiplet of
$G_F$, it may point in different directions in flavor space at
different $y$; this is a situation which arises when multiple
source branes are present. At tree level the only flavor breaking is
given by $\phi(0)$, whereas at the loop level the flavor breaking of
$\phi(y)$ is also probed. We say that additional flavor breaking is
``sniffed'' in the bulk from points $y\ne0$.

In addition to quantum corrections to $<\!\!\phi^p\!\!>$, there can be 
classical corrections as well. By integrating out a $\phi$ field which 
interacts in the bulk, we generate local operators as we illustrate in 
figure \ref{fig:nlints3}(b). In this particular example, given
operators $\phi(x_1) O_1(x_1)$ and $\phi(x_2) O_2(x_2)$, we generate a
local operator $\phi^2(x) O_1 O_2(x)$.  Absent bulk corrections, this
operator would have a
coefficient $<\!\!\phi\!\!>^2$, but sniffing contributions can change
this. Although both $\phi$-legs on our wall are evaluated at the same
point in spacetime, this is not a quantum effect and does not receive
the same loop suppression as in Fig. \ref{fig:nlints3}(a).

\subsection{Spatial derivatives of the flavon field}

The bulk flavon fields have $y$ dependent profiles, and,
because Lorentz invariance is violated
in directions perpendicular to our brane, one might imagine that 
wherever a $\chi$ field appears in the Lagrangian of our brane, we
could just as easily write $(a_n \partial_n +a_{mn}\partial_m
\partial_n+...)\chi$,
with no need to contract indices of extra dimensional 
derivatives \cite{BDN}.
Unless the $\chi$ mass were
much smaller than the scale $\Lambda$, the derivative terms would
not be strongly suppressed.  Nor would $\chi$ and its derivatives 
necessarily be proportional to each other, since the field can have sources
on several branes, potentially leading to a variety of troublesome
flavor changing effects. 
However, if Lorentz invariance is broken {\em spontaneously} (as
is the case if standard model fields are localized on a D-brane, 
for instance), only certain derivative terms are allowed.  In 
the low energy effective theory, we simply have SM fields localized 
to our brane, the bulk
$\chi$ field,  and the goldstones of spontaneously broken translational 
invariance $Y^m$, which give the position of our brane in the extra 
dimensions \cite{S1}. Thus, all terms involving
derivatives of a single power of $\chi$ must feature either $\Box^j
\chi$ for some integer $j$, or a brane tension-suppressed coupling to the
goldstone.  For instance, we can have terms like
\begin{equation}
\partial_\mu Y^m ({\overline Q}\, \partial_m \chi\, {\overline \sigma}^\mu 
Q) \hspace{.3in} {\rm or} \hspace{.3in} Q\Box \chi D^c H,
\end{equation}
but not something like
\begin{equation}
Q\,\partial_{m} \chi\, D^c H,
\end{equation}
because extra-dimensional derivatives with uncontracted indices amount to 
{\em explicit} breaking of Lorentz invariance.

Of course, the localized fields have finite profiles in the
bulk, and we can contract derivatives of $\chi$ with derivatives
of wall fields in the full extra-dimensional theory. 
However, the effective field theory argument just given
indicates that in our wall's low energy theory only terms
involving $\Box^j \chi$ will be generated.  It is straightforward to
see how this comes about explicitly from a microscopic description. 
 Let us label the $(4+n)$ dimensional 
spacetime coordinates as $(x^\mu, y^m)$, where $x^\mu$ are
the 4D coordinates on our brane, and $y^m$ are the coordinates of the
extra $n$ dimensions.  We will use an index $K = (\mu,m)$ that runs over
all $(4+n)$ dimensions.  Consider the Lorentz invariant term
\begin{equation}
{1 \over \Lambda^{n+3}} \int \! d^4 \! x \int \! d^n \! y \,
 (\partial_K Q^{(4+n)}) 
(\partial^K \chi^{(4+n)}) {D^c}^{(4+n)} H^{(4+n)}.
\label{eq:LI}
\end{equation}
We assign the standard model fields $\psi^{(4+n)}$ Gaussian 
profiles in the extra dimensions, so that their relation to the
canonically normalized fields in 4D is
\begin{equation}
\psi^{(4+n)} (x,y) = \left( {2 \Lambda^2\over \pi}\right)^{n/4} \psi^{(4)} (x) 
e^{-\Lambda^2 |y|^2},
\end{equation} 
while the $x$-independent bulk flavon VEV is given by
\begin{equation}
\chi^{(4+n)} (x,y) = \Lambda^{n/2} \chi^{(4)} (y).
\end{equation}
In terms of the canonically normalized fields, (\ref{eq:LI}) becomes
\begin{equation}
-{\Lambda^{n-1}} \int \! d^4 \! x \, Q^{(4)}(x)\left( \int \!d^n \! y \, 
e^{-3 \Lambda |y|^2}y^m \, \partial^m \! \chi^{(4)}(y) \right){D^c}^{(4)}(x)
H^{(4)}(x),
\end{equation}
where we have neglected factors of $\pi$ and 2.  After integration by
parts, the piece involving $\chi$ becomes
\begin{equation}
\int \!d^n \! y \, (6 \Lambda^2 |y|^2 - n) \chi^{(4)} (y)e^{-3 \Lambda |y|^2}.
\end{equation}
This is of the form $\int \! d^n \! y \, f(y) e^{-\alpha |y|^2}$,
which is equivalent to 
\begin{equation}
\int \! d^n \! q \, {\tilde f}(q) e^{-|q|^2 / \alpha}
=\left. (e^{-\Box /\alpha}f)\right|_{y=0},
\end{equation}
where $\tilde{f} (q)$ is the Fourier transform of $f(y)$.
In our case, we have $f(y)=\chi^{(4)}(y) (a+b|y|^2)$, with $a$ and 
$b$ real, which satisfies
\begin{equation}
\left.e^{-\Box /\alpha}(\chi^{(4)}(a+b|y|^2))\right|_{y=0}
=\sum_{j=0}^\infty c_j \left.\Box^j \chi^{(4)}\right|_{y=0},
\end{equation}
with real coefficients $c_j$.  Although there can be many $\chi$ derivatives
that appear in the Yukawa interaction, we see that they are all 
proportional to 
$\Box^j \chi$ for some integer $j$ and thus, by the equations of
motion, all proportional to $\chi$ itself. 

In contrast, an operator like
\begin{equation}
 \partial_m \chi \partial^m \chi O_{SM}(x),
\label{eq:deriv2}
\end{equation}
where $O_{SM}$ is an operator of standard model fields, cannot be
brought into a form involving only $\Box^j \chi_d$. The
presence of these operators has model dependent effects which we will
discuss in later sections.

\subsection{Harmless flavon exchange}

As discussed in section 2.4, breaking flavor symmetries at low $\Lambda$ 
in four dimensions generates harmful flavor-changing
operators through the
exchange of flavons, due to the \, smallness of the flavor-breaking
scales relative to the
fundamental scale. 
In sharp contrast, there is no reason in extra dimensional theories to
expect that the bulk flavon masses are closely related to the sizes of the 
Yukawa couplings, as these small parameters are no longer ratios
of mass scales. However, even if the bulk fields {\em were} very light,
the harmful operators still receive no subsequent enhancement. This is 
due to the IR softness of bulk propagators in extra
dimensions. Returning to the example of section
2.4, let us suppose that $\phi_{12}$ lives in $p$ extra
dimensions. Then, the coefficient of the induced 4-fermi operator is
(working in units with $M_*=1$) 
\begin{equation}
v^2\int \frac{d^p \kappa}{(2 \pi)^p} \frac{1}{\kappa^2 + m^2_{\phi}}.
\end{equation}
Note that we integrate over the extra dimensional momenta $\kappa$ since 
this momentum is not conserved. The important point is that for $p>2$, 
this integral is dominated in the $UV$ and is insensitive to $m_{\phi}$!
Therefore, the generated operator is {\it not} enhanced by $1/m_{\phi}$ factor.
In fact, the dominant contribution is $m_\phi$ independent and generates 
a flavor-symmetric operator suppressed by powers of $M_*$ and additional
loop factors. Sub-dominant contributions need
{\it not} respect the flavor symmetry, but are small enough; the 
leading corrections go as 
\begin{equation}
L(p) \log m_{\phi}^2, \,\,p=2; \hspace{.3in} L(p) m_{\phi}, \,\, p=3;\hspace{.3in}  L(p)  m_{\phi}^2, \,\, p=4,5,\cdots
\end{equation}
where 
\begin{equation}
L(p) = \frac{1}{p 2^{p-1} \pi^{p/2} \Gamma(p/2)}
\end{equation}
is a loop factor.

We can also return to the example of our toy $U(2)$ theory discussed in
section 2.5. The tree-level exchange of a {\em bulk} $A$ field
produces the operator
\begin{equation}
L(n) \frac {H^* H (Q_a D^c_b)(\bar{Q}^a \bar{D}^{cb})}{M_*^4}.
\end{equation}
There is no inverse dependence on the $A$ mass at all, and we
can therefore tolerate $M_* \sim 1-10$ TeV, roughly three orders of
magnitude below the bound on $\Lambda$ in the 4D case.

\section{A $U(3)^5$ theory in extra dimensions with 3 source branes}

Having introduced the shining of flavor breaking from
distant branes, we now discuss the construction of $U(3)^5$ models in
extra dimensions.   In the models we will describe, 
$\chi_u$ and $\chi_d$ (which in this context are bulk fields) 
will be the only flavons that couple directly to the standard model 
fields of our 4D universe, just as in minimal $U(3)^5$ in 4D.  
The authors of \cite{AD} have applied
the shining framework described in section 3 to the case of $U(3)^5$.
In their picture, $\chi_u$ ($\chi_d$) couples to nine source 
fields $\phi^{u, ij}$ ($\phi^{d, ij}$).  Each of these source
fields transforms as $\chi_u$ ($\chi_d$) and is localized on
its own distinct brane.  The sources acquire VEVs of the form
\begin{equation}
<\phi^{u, ij}_{kl}>\sim <\phi^{d, ij}_{kl}>\sim \Lambda \delta^i_k
\delta^j_l,
\label{eq:vev}
\end{equation}  
that is, each of the nine $\phi^d$ sources essentially shines a 
single element of the down quark mass matrix, and similarly for the up
sector.  The magnitude of each source is taken to be roughly
$\Lambda$, but large fermion mass ratios are still possible by 
requiring some source branes to be closer to our brane 
than others.  This represents a significant improvement over the 
minimal case in 4D:  only a single $\chi_d$ and a single 
$\chi_u$ appear in the Yukawa interactions, and yet a simple 
explanation for the hierarchical nature of the fermion masses is achieved.

This picture is far from complete, however.    
The most serious deficiency is that no understanding is provided
of why $V_{CKM} \sim I$.  Related to this is the fact that the
VEVs of (\ref{eq:vev}) do not comprise a justifiable starting
point, as we will now argue.  To avoid problems with goldstone bosons, 
we work with a large
discrete subgroup of $U(3)^5$ rather than with $U(3)^5$ itself
(because the breaking is order unity, we avoid the light pseudo-goldstone
bosons that appear in the 4D case).  The
directions of the eighteen $\phi^u$ and $\phi^d$ VEVs are thus fixed
in various directions that do not depend on  bulk dynamics.
The important point is that there is no reason for the direction 
of a source on
one brane to be related in any particular way to the direction of a 
source on another.  This reasoning argues against the
arrangement of VEVs in (\ref{eq:vev}), and more generally, it tells
us that we should expect order unity  CKM mixing angles if all the sources
are on separate branes.  Suppose, for example, that a $\phi^u$
localized on one nearby brane breaks 
$U(3)_Q \times U(3)_{u^c} \rightarrow U(2)_Q \times U(2)_{u^c}$, while 
on another brane, a $\phi^d$ independently breaks $U(3)_Q \times
U(3)_{d^c} \rightarrow U(2)_Q \times U(2)_{d^c}$.
We imagine that these branes are the ones nearest us, so that these 
$\phi$'s shine the leading order
contributions to the quark mass matrices.  The $U(3)_Q \times
U(3)_{d^c}$ symmetry allows us to take
\begin{equation}
\phi^d = \pmatrix{0&0&0 \cr 0&0&0 \cr 0&0&v},
\label{eq:diag}
\end{equation} 
and then, using the $U(3)_{u^c}$ symmetry, we can write write
\begin{equation}
\phi^u = M\pmatrix{0&0&0 \cr 0&0&0 \cr 0&0&v'},
\label{eq:nondiag}
\end{equation} 
where the form of $M \in U(3)$ is fixed by the explicit breaking.  
The point is that there is no reason for $\phi^u$ and $\phi^d$ to
choose the same unbroken $U(2)_Q$, and there is not in general 
a basis in which both 
$\phi^u$ and $\phi^d$ are diagonal, because the $U(3)_Q$ freedom is
used up entirely in diagonalizing either one or the other.
Generically, we expect the (23) entry of $\phi^u$ to be
roughly as large as its (33) entry\footnote{Using the residual $U(2)_Q$
symmetry respected by $\phi^d$, the (13) entry can be made to
vanish.}, and since the leading order 
form of $\chi_{u(d)}$ on our brane is simply proportional to  
$\phi^u$ ($\phi^d$), we should expect a large CKM mixing angle, 
contrary to what is observed.
\subsection{A complete $U(3)^5$ model}
We now describe a model that retains
the successes of the picture just described, but which in addition
predicts small mixing angles.  The model is remarkably simple.  We
assume the existence of a series of source branes, each of which has
localized on it a triplet under $U(3)_Q$, a triplet under $U(3)_{u^c}$,
and a triplet under $U(3)_{d^c}$.  Nothing special distinguishes any
of the source branes - we will even assume for simplicity that they are identical copies of each
other - except that they are located at different distances
from our brane.  The three triplets on each brane acquire VEVs near
the fundamental scale and act as sources for bulk flavons $\chi_u$ and
$\chi_d$.  We again regard the true flavor group as a
large discrete subgroup of $U(3)^5$ to avoid goldstones, so the
potential for each triplet features a discrete series, rather than
a continuum, of minima. Each triplet's VEV is stuck at one of these
minima, unable to tunnel from one to another.  Moreover, the
directions chosen by the sources on one brane
are not related in any particular way to the directions chosen
on a different brane.  What we have, effectively, is
explicit breaking on each source brane, with the triplets getting fixed,
complex VEVs that point in uncorrelated directions.

Let us work out the implications of this simple scenario.  On the brane 
nearest ours, the triplet sources acquire VEVs that, if we exploit our 
$U(3)^3$ freedom, we can write as
\begin{equation}
T_Q^1 = \pmatrix{0\cr 0\cr v_Q}, \hspace{.3in}
T_u^1 = \pmatrix{0\cr 0\cr v_u}, \hspace{.1in}{\rm and}\hspace{.1in} 
T_d^1 = \pmatrix{0\cr 0\cr v_d},
\label{eq:brane1}
\end{equation}
with $v_Q$, $v_u$, and $v_d$ real and not much smaller than $M_*$. 
In fact, these sources could even be localized on {\it our} brane. 
The bulk flavons are shined by the triplet sources
due to the brane interactions\footnote{ If the sources are on our 
brane, the third generation quarks acquire mass from direct 
couplings to the triplets.}
\begin{equation}
T_Q^1 \chi_u T_u^1 \hspace{.3in} {\rm and}\hspace{.3in}
T_Q^1 \chi_d T_d^1. 
\end{equation}
Consider, for the moment, the extreme case in which this brane is 
by far the closest one to our ours - the closest by so much that, on
our wall, we can ignore contributions to flavon VEVs coming from all
other sources.  
In contrast to the example that led to the alignment problem of
equations (\ref{eq:diag}) and (\ref{eq:nondiag}), both $\chi_u$ and 
$\chi_d$ are shined from the
same nearby brane, and simultaneously take the form
\begin{equation}
\chi_{u,d} \propto \pmatrix{0&0&0 \cr 0&0&0 \cr 0&0&A_{u,d}}.
\end{equation} 
Now let us consider the additional contributions to $\chi_{u,d}$ that
are shined from more distant sources.  The VEVs of (\ref{eq:brane1})
respect a residual $U(2)^3$ symmetry that can be used to write
the sources on the second nearest brane as
\begin{equation}
T_Q^2 =v_Q \pmatrix{0\cr \sin\theta_Q \cr \cos\theta_Q e^{i\alpha_Q}}, \hspace{.15in}
T_u^2 =v_u \pmatrix{0\cr \sin\theta_u \cr \cos\theta_u e^{i\alpha_u}}, \hspace{.07in}{\rm and}\hspace{.07in} 
T_d^2 =v_d \pmatrix{0\cr \sin\theta_d \cr \cos\theta_d e^{i\alpha_d}},
\label{eq:brane2}
\end{equation}
where we assume for simplicity that $T^{\dagger} T$ is the same in the
various discrete minima. 
After we include the effects of the shining interactions
\begin{equation}
T_Q^2 \chi_u T_u^2 \hspace{.3in} {\rm and}\hspace{.3in}
T_Q^2 \chi_d T_d^2, 
\end{equation}
the flavon VEVs on our brane take the form
\begin{equation}
\chi_{u,d} \propto \pmatrix{0 & 0 & 0 \cr 0
  & \epsilon & \epsilon \cr 
0 & \epsilon & A }_{u,d}, 
\end{equation}
with $\epsilon_{u,d} \ll A_{u,d}$.  At this stage, the VEVs in 
(\ref{eq:brane1}) and (\ref{eq:brane2})
still admit a $U(1)^3$ symmetry that can be used to write the sources
on the third brane as
\begin{equation}
T_Q^3 =v_Q \pmatrix{{\rm s}\phi_Q \cr {\rm c}\phi_Q\, {\rm s}\rho_Q\, e^{i\beta_Q} \cr {\rm c}\phi_Q\, {\rm c}\rho_Q\, e^{i\gamma_Q}}, \hspace{.1in}
T_u^3 =v_u \pmatrix{{\rm s}\phi_u \cr {\rm c}\phi_u\, {\rm s}\rho_u\, e^{i\beta_u} \cr {\rm c}\phi_u\, {\rm c}\rho_u\, e^{i\gamma_u}}, \hspace{.1in} 
T_d^3 =v_d \pmatrix{{\rm s}\phi_d \cr {\rm c}\phi_d\, {\rm s}\rho_d\, e^{i\beta_d}
  \cr {\rm c}\phi_d\, {\rm c}\rho_d\, e^{i\gamma_d}}.
\label{eq:brane3}
\end{equation}
Including the effects of this brane, we arrive at the Yukawa texture
\begin{equation}
\lambda_{u,d} \sim \pmatrix
{\epsilon' &\epsilon'&\epsilon'\cr \epsilon'
  & \epsilon &\epsilon\cr 
\epsilon'& \epsilon & A}_{u,d},  
\label{eq:texture}
\end{equation}
which features both a hierarchy of eigenvalues and small mixing
angles.  

This model features a simple symmetry breaking pattern.  If we include
only the nearest brane, its sources break  $U(3)^3 \rightarrow
U(2)^3$.  Bringing the second brane into the picture then 
breaks $U(2)^3 \rightarrow U(1)^3$.  Finally, moving the third brane 
into place breaks $U(1)^3 \rightarrow$ nothing.  Note that this 
breaking pattern is not
put in by hand, but rather follows inevitably from the fact that
the sources transform as triplets and acquire fixed VEVs pointing
in random different directions.  Note also that we work with three
source branes only because this is the minimal set required to break
$U(3)^3$ entirely.  Given that at least three exist, the success of
our picture is insensitive to how many branes there are in all.  
Additional branes, being further away, will give small contributions
to the Yukawa couplings, leaving the texture of equation
(\ref{eq:texture}) unchanged.

The Yukawa texture suggests the approximate relations 
$|V_{ij}| \sim {m_{d_i}}/ {m_{d_j}}$, 
which work reasonably well for all mixing angles except
for $\theta_c$.  The fact that $\theta_c$ naively comes out to small
is not a serious problem, because the entries of (\ref{eq:texture})
come with unknown coefficients of order 1 due to the unknown angles 
$\theta$, $\phi$, $etc.$, that appear in equations (\ref{eq:brane2}) and
(\ref{eq:brane3}):  by taking $\tan \phi_d \sim {1 \over 5} \sin
\rho_d$ we obtain the correct
size $\theta_c \sim 1/4$.  Somewhat surprisingly, the more
closely {\em aligned} the source triplets on the second and third
branes are, the {\em larger} $\theta_c$ is.  In sections 6 and 7 we consider
different models that accomodate $\theta_c$
more easily.  

It is advantageous to give the flavons
slightly different masses, with $m_{\chi_u}>m_{\chi_d}$, 
to explain why the mass hierarchies are stronger for the up quarks than for
the down quarks.  Doing so makes all the more pressing the question of
why $m_t>m_b$, given that these masses are essentially shined from the 
same brane.  Having
two Higgs doublets with large $\tan \beta$ leads to flavor changing
problems, as we will see below.  A simple alternative that leads to no 
phenomenological difficulties is to have a $\sim 1/60$
suppression of $v_d$ relative to $v_u$.  Even irrespective
of flavor changing issues, this approach may be more appealing
than a non-SUSY large $\tan \beta$ scenario, because here the different-sized
VEVs are given to two fields, $T_u$ and $T_d$, that transform entirely
differently under the flavor symmetry.  In contrast, if we have 
two Higgs doublets, $H_d$ and $\tilde H_u$ transform identically under
the gauge symmetry and are both flavor singlets, so it is  especially
difficult to understand how one is chosen to have a much larger VEV
than the other.  A more interesting approach to understanding $m_b/m_t$
will be described in section 7.  The details of how the present  $U(3)^5$ model can give
realistic fermion masses and mixings are important but should not 
obscure the central point:  having source triplets with
uncorrelated VEVs leads {\em automatically} to a CKM matrix
with small mixing angles.

Another attractive feature of this model is that it
violates CP spontaneously, as the VEVs of 
equations (\ref{eq:brane2}) and (\ref{eq:brane3}) give the
off-diagonal elements of $\lambda$ order unity phases.  
If we impose CP
as a symmetry of the underlying theory, this model exhibits the same 
solution to the EDM problem described in section 2.6 in the context
of $U(3)^5$ in 4D.  
\subsection{Flavor-changing from the bulk}
The only question is whether there are additional 
challenges in suppressing  dangerous operators, now that we are working
in extra dimensions.  As discussed in section 3, physics in our 4D
universe can be sensitive not only to the values of 
$\chi_u$ and $\chi_d$ on our brane, but also to their values away from
our brane, due to ``sniffing'' effects.  For
instance, if we have bulk couplings $\int \! d^n \!y \, 
Tr(\chi_u^\dagger \chi_u \chi_u^\dagger \chi_u)$, and  $\int \! d^n \!y \, 
(Tr(\chi_u^\dagger \chi_u))^2$, then in the up
quark EDM operator we can replace the matrix $\langle \chi_u (y=0)\rangle_{ij}$ with 
\begin{equation}
{\langle \chi_u^\dagger
(0)\rangle_{mn} \over 16\pi^2} \int \! d^n \!y \, \langle {\chi_u}_{im} 
{\chi_u}_{nj} (y) \rangle \Delta^2(y),
\label{eq:oneloop}
\end{equation}
or with 
\begin{equation}
{1 \over (16\pi^2)^2} \int \! d^n \!y \, \langle \chi (y) \rangle_{ij} \Delta^3(y),
\end{equation}
where we have included loop factors from integrating over 4D 
momenta.  Diagrams representing these contributions are shown in
Figs. \ref{fig:nlints3}a and \ref{fig:nlints2}b, respectively.  
Because they do not have the same flavor structure as $\chi_u$, we
need to check that these contributions are not problematic.  The largest
contribution to the 1-1 entry comes from the piece of
(\ref{eq:oneloop}) proportional to $\lambda_t$,
\begin{equation}
{\lambda_t \over 16\pi^2}\int \! d^n \!y \,\langle {\chi_u}_{13} 
{\chi_u}_{31} (y)\rangle \Delta^2(y).
\end{equation}
$\Delta(y)$ is largest near our brane ($y \sim 0$), but in this region
$\langle {\chi_u}_{13} {\chi_u}_{31} (y)\rangle$ nearly vanishes in the mass
diagonal basis.  On the other hand,$\langle {\chi_u}_{13}
{\chi_u}_{31} (y)
\rangle$
is largest near the third most distant brane, located at $y_3$.  Using
the short distance form for the propagator to evaluate 
$\langle {\chi_u}_{13} {\chi_u}_{31} (y) \rangle$ in this region, we get a
contribution of roughly
\begin{equation}
{\lambda_t \over 16\pi^2} \lambda_u^2 {1\over dS_n} f(m_\chi ),
\end{equation}
where $f(m_\chi ) \sim ({1 \over m_\chi})^{(4-n)}$, $\mbox{Log}(
  m_\chi)$, and 1 for $n=$ 2 or 3, $n=4$, and $n>4$ respectively, and
where $dS_n$ is the surface area of the unit sphere in $n$ dimensions.  The
potential mild enhancement from $f(m_\chi)$ cannot nearly compensate
for the extra factor of ${\lambda_u \over 16\pi^2}$ relative to what 
we have for $\langle \chi_u(0) \rangle_{11}$, so this contribution is harmless.  
Sniffing contributions to the down quark EDM are similarly suppressed.

Sniffed versions of
\begin{equation}
(Q \chi_d D^c)(Q \chi_d D^c)^\dagger \hspace{.4in} {\rm and}
\hspace{.4in}(Q \chi_d D^c H)(Q \chi_d D^c H)^\dagger
\end{equation}
yield $\Delta$S $=2$ operators with  coefficients of approximate size
\begin{equation}
{1 \over 16\pi^2} \int \! d^n \!y \, \langle {\chi_d}_{12} 
{\chi_d}^*_{21} (y)\rangle \Delta^2(y),
\end{equation}
and
\begin{equation}
v^2 \int \! d^n \!y \, \langle {\chi_d}_{12} 
{\chi_d}^*_{21} (y)\rangle \Delta^2(y),
\end{equation}
respectively.  Again concentrating on the region around $y_3$,
we estimate the integrals as roughly $(\lambda_s
\theta_c)^2 {1\over dS_n} f({m_\chi})$.  If the coefficient were
simply $(\lambda_s \theta_c)^2$, $\epsilon_K$ would require
$\Lambda>7$ TeV.  Since these contributions are further suppressed by
either a loop factor or by $v^2$, they are safe.

A different challenge posed by the extra dimensions
involves the bulk flavon derivatives described in section 3.3.  If we
allowed all flavor invariant terms with extra dimensional derivatives 
acting on $\chi_d$, then in the basis that diagonalized and made real 
the down quark Yukawa interaction 
\begin{equation}
{Q} \left( (1+a_n \partial_n +a_{mn}\partial_m \partial_n+...)\chi_d\right)
D^c H,
\label{eq:Yukd}
\end{equation}
the EDM operator 
\begin{equation}
F_{\mu \nu} Q \left( (1+b_n \partial_n +b_{mn}\partial_m
  \partial_n+...)
\chi_d\right) \sigma^{\mu \nu} D^c H
\label{eq:EDMd}
\end{equation}
would in general be complex, leading to the familiar EDM bound 
$\Lambda>$ 40 TeV. However, as discussed in section 3.3, 
the only drivatives of $\chi_d$ allowed in equations (\ref{eq:Yukd})
and (\ref{eq:EDMd}) are those of the form $\Box^j \chi_d = m^{2j}
\chi_d$, provided Lorentz invariance is broken spontaneously.  In this
case derivative terms are harmless as far as EDM's are concerned.
On the other hand, the operator
\begin{equation}
{1 \over \Lambda^2}\left( {Q} {\partial_K \chi_d \over \Lambda^2}
D^c\right) \left( {Q}{\partial^K \chi_d \over \Lambda^2}
D^c\right) ^\dagger
\label{eq:deriv}
\end{equation}
cannot be brought into a form involving only $\Box^j \chi_d$.
Because $\partial_m \chi_d$ is not in general proportional to $\chi_d$, the 
most pessimistic view is then that $\epsilon_K$ forces us to take 
$\Lambda>7$ TeV.  (In fact, at this point it becomes clear why using
large $\tan \beta$ to explain $m_t>m_b$ is disasterous: the $\Delta$S $=2$
piece of (\ref{eq:deriv}) has a coefficient that is proportional to
$\tan^2 \beta$).
Note, however, that the $\Lambda>7$ TeV interpretation assumes that the
derivative terms are entirely unsuppressed:  if $m_{\chi}=\Lambda 
/S$,
for instance, then the bound is reduced by a factor of $S$.  Also, it is
conceivable that $\partial_a \chi_d$ {\em is} nearly
proportional to $\chi_d$.  For example, if the source branes lie along
along the same direction from ours, then in the case of three extra
dimensions, the derivative contributions that are not proportional to
$\chi_d$ are suppressed by factors of $1/(\Lambda r_i)$
relative to the leading non-derivative contribution, where $r_i$ are 
the various distances of the source branes from our brane.
$K-{\overline K}$ mixing is most sensitive to contributions shined
from branes responsible for the light quark masses.  
Taking $(mr) \sim 5$ for these branes, and $m \sim \Lambda /3$, we
find that the bound on $\Lambda$ is reduced by a factor $\sim 15$.
The general point is that bounds 
derived by considering  terms involving flavon derivatives are softer than  
those obtained from operators without derivatives, as they are more
sensitive to the flavon mass, and to the details of the brane configuration.  

We have seen that $U(3)^5 \times$CP models with triplet sources are 
generically safe for $\Lambda \approx$ 5 TeV, provided that Lorentz
invariance is broken spontaneously.  
Next we will show that specific models can be safe at this scale 
without qualification.  
In particular, we present what we
consider the simplest specific realization of our $U(3)^5 \times$CP
scenario, and find that
regardless of how Lorentz invariance is violated, and regardless of
whether $m_{\chi}$ is suppressed relative to $\Lambda$, both flavon
derivative and ``sniffing'' effects are harmless.

\section{A concrete realization of $U(3)^5$}
Here, we will consider a concrete arrangement of branes in our $U(3)^5$ 
scenario. The arrangement is very simple and furthermore allows analytic 
calculation of FCNC effects. We will see that 
the potential flavor-changing effects are very suppressed by this particular
set-up; for instance all $\chi$ derivatives are exactly aligned  
with $\chi$. 

We imagine that even though there are $n \geq 2$ extra spatial
dimensions, flavor is associated with only one of them, which we
parametrize by $y$.  Our 3-brane and several source 3-branes are
taken
to lie in a 4-brane described by ($x$, $0 \leq y \leq L$),
where we compactify on an interval $[0,L]$ (rather than a circle) of
moderately large size, $L \sim 10 M_*^{-1}$.  The 3-branes are 
spread out roughly
evenly in the space available to them and are then naturally 
spaced between $\sim 1-10$ times $M_*^{-1}$, so that the question of what
determines the sizes of the inter-brane 
separations is to some extent obviated.  The flavon $\chi$ is taken to
propogate only on this 4-brane. 

In one infinite extra dimension, the $\chi$ propagator is just
$e^{-m|y-y'|}$. When the dimension is compactified, the propagator depends on the boundary 
conditions. We will impose the conditions $\chi=0$ at the boundaries 
of the interval,
so that the propagator on the strip from $y_1$ to $y_2$ is 

\begin{equation}
\Delta(y_1,y_2) = \left(\theta(y_1 - y_2) 
\frac{\mbox{sinh} [m(L - y_1)]}{\mbox{sinh}[mL]} 
\mbox{sinh} [my_2] + \theta(y_2 - y_1) \frac{\mbox{sinh} [my_1]}{\mbox{sinh}[m L]} 
\mbox{sinh} [m(L - y_2)] \right)
\end{equation}
Note that this goes to $e^{-m|y_1 - y_2|}$ when 
$L \rightarrow \infty$, as it should. 
The classical profile for $\chi$ is then 
\begin{equation}
\chi_{cl}(y) = \sum_i \chi_i \Delta(y_i,y),
\end{equation}
where
\begin{equation}
\chi_{i(a \alpha)} = T_{Lia} T_{Ri\alpha}
\end{equation}
is the source for $\chi$ shone from the $i$'$th$ wall. 
Suppose that our 3-brane, located at $y=y_*$, is positioned to the left of 
all the other branes on the strip, i.e. $ y_*<y_3$ where $y_3$ 
is the location of the 
nearest wall. Then, for all $y<y_3$ the 
profile of $\chi$ is 
\begin{equation}
\chi_{cl}(y) = \bar{\chi} \, \mbox{sinh} [m y] ,
\end{equation}
where
\begin{equation}
\bar{\chi} = 
\sum_i \chi_i \frac{\mbox{sinh}[m(L - y_i)]}{\mbox{sinh}[mL]}.
\end{equation}  
Note that the $\chi_{cl}(y)$ at different values of $y<y_3$,
are all proportional to the same matrix, and so all derivatives of $\chi_{cl}$ 
evaluated on our wall are diagonal in the same basis as $\chi_{cl}$ itself. 
Therefore, even allowing for heavy $m_{\chi} \sim \Lambda$ and explicit 
Lorentz violation in the extra dimension, there are no problems with 
derivative terms. In the absence of interactions in the bulk, the 
FCNC analysis is identical to the standard $U(3)^5$ spurion analysis. 

In four dimensions, $U(3)^{5}$ with minimal flavons leads to exact lepton 
flavor conservation. As we have discussed, 
in higher dimensions, derivative 
operators have the potential to induce some level of flavor changing, 
but in this simple realization of $U(3)^{5}$, where $\partial_{m} \chi \propto 
\chi$, one might expect again that flavor changing is 
absent. This intuition is incorrect, as sniffing effects 
give us sensitivity to the value of $\chi$ in the bulk, and thus to 
regions where it is not diagonal in the mass basis. However, it is 
easy to see that these effects are highly suppressed.

We can illustrate this by considering the process $\mu 
\rightarrow 3e$, which occurs due to the presence of the operator 
\begin{equation}
E^{c}_{k} \chi^{kl} \chi^{\dagger}_{lm} \overline E^{c \> m} E^{c}_{n} 
\overline E^{c \> n}.
\end{equation}
In addition to the spurion contribution, which 
is diagonal in the flavor basis, we have the sniffed contribution
\begin{equation}
<\chi \chi^{\dagger}> = \int dy d^4 x \chi_{cl} \chi^{\dagger}_{cl} (y)
\Delta(y,x)^2.
\label{eq:linsniff1}
\end{equation}
If we assume a brane geometry where our 
brane is at $y=0$, and the $\mu$ and $e$ branes are at positions
 $0< y_\mu < y_e$, the sniffed 
contributions from the region $y<y_\mu$ will all be proportional to
$\chi$. Thus, the first flavor changing piece comes in the region
$y_\mu < y<y_e$. We calculate the coefficient of the operator
 to be

\begin{equation}
<\chi \chi^\dagger>_{FC}^{12} \simeq 
{\lambda_e \lambda_\mu m^{3} \over 32 \pi^2 (\mbox{Log}\lambda_{\mu})^{2} }
\approx 10^{-13},
\label{eq:linsniff2}
\end{equation}
which gives a completely unobservable rate for $\mu \to 3 e$. Thus, while 
sniffing does allow for flavor violations even in a $U(3)^5$ theory (where
naively lepton flavor is conserved!), they are easily small enough to be 
harmless.



\section{Smaller flavor symmetries}

In the previous two sections we developed extra-dimensional
models of flavor with low $\Lambda$ and $G_F=U(3)^5$.  One might wonder 
how difficult it is to work instead with smaller flavor groups.  For
instance, we have already seen that taking the symmetry to be
$U(2)$ is problematic:  the extra dimensions
alleviate the flavon exchange problem, but the $U(2)$-invariant,
non-renormalizable operator of equation (\ref{eq:badop}) forces
$\Lambda> 10^5$ TeV.  Of course, we could hope
that this operator is simply not generated by the underlying
theory, but if one wants to assume that all invariant operators
are present, then we need a larger symmetry.  Here we
adopt the group $U(2)^3$ and consider the quark sector alone.
We again take CP to be a symmetry of the underlying theory in hopes
of evading the EDM bound.

With this choice of flavor symmetry both $Q_3 D^c_3 H$ and $Q_3 U^c
{\tilde H}$ are flavor singlets, so to explain $m_b \gg m_t$, we
might require two Higgs doublets with large $\tan\beta$.
Unfortunately, as we have already seen, unless a single flavon
multiplet is responsible for the elements of $\lambda_d$ involving
the light generations (which will not be the case for $U(2)^3$), then
we expect to get the operator
\begin{equation}
{(\lambda_s \theta_c \tan\beta)^2 \over \Lambda^2}(Q_2
D^c_1)({\overline Q}_1{\overline D}^c_2),
\end{equation}
forcing $\Lambda>400$ TeV.  Thus, we instead use a single Higgs 
but enlarge the flavor symmetry to include an
extra $U(1)$ factor under which only $D^c_3$ is charged.  We introduce
a bulk flavon $\theta$, whose charge under this $U(1)$ is opposite
that of $D^c_3$, and take the VEV of $\theta$ on our wall to be $\sim
1 /60$.

Next we introduce another bulk flavon, $\phi$, a
doublet under $U(2)_Q$.  We can choose a basis in which its VEV
on our wall is 
\begin{equation}
\phi =\pmatrix{0 \cr v}, 
\end{equation}
with $v$ real, and to yield a reasonable $V_{cb}$ we take $v \sim
1/30$.  Our picture for
the symmetry breaking that gives masses to the light generations is
designed to preserve certain features of the standard $U(2)$ fermion mass
texture, in particular the relation $\theta_c \approx \sqrt{m_d/m_s}$.
We imagine that on a distant brane the subgroup $U(2)_Q \times
U(2)_{d^c}$ is broken down to $U(2)$ by a source that shines the bulk
flavon $\chi_d$, transforming as (2, ${\overline 2}$) under this subgroup.
Similarly, from a different brane we have $U(2)_Q \times
U(2)_{u^c} \rightarrow U(2)$ breaking transmitted by $\chi_u$, a (2,
${\overline 2}$) under {\em this} subgroup (note that we do not expect
the two walls to preserve the same $U(2)$).  Finally, we imagine that 
on our brane both $U(2)_{d^c}$ and $U(2)_{u^c}$ are broken {\em primordially}
to their $SU(2)$ subgroups.  

What does this assortment of breakings say about the Yukawa matrices?
The flavor symmetry allows us the freedom to choose convenient forms
for the $\chi_u$ and $\chi_d$ VEVs, but because of the primordial
breaking on our brane, and because of the freedom already used to fix 
the form of $\phi$, we are only allowed arbitrary
$SU(2)_{u^c} \times SU(2)_{d^c} \times U(1)_Q$ transformations,
where the $U(1)_Q$ acts on $Q_1$ alone.  
These transformations allow us to take
\begin{equation}
\chi_d = v_d \pmatrix{ 1 & 0 \cr 0 & 1} \hspace{.5in} {\rm and}
\hspace{.5in}
\chi_u = v_u \pmatrix{ e^{-i\delta} & 0 \cr 0 & 1}
\end{equation}
on our wall, with both $v_u$ and $v_d$ real.  We are stuck with a phase in
$\chi_u$; this is the origin of CP violation in the model.  The
leading order couplings of these flavons to the quarks are
\begin{equation}
Q_L {{{\overline {\chi}}_d}^L}_l \epsilon^{lm} {D^c}_m \hspace{.5in} {\rm and}
\hspace{.5in}
Q_L {{{\overline {\chi}}_u}^L}_l \epsilon^{lm} {U^c}_m,
\end{equation}
so that at this stage the Yukawa textures are
\begin{equation}
\lambda_D = \pmatrix{ 0 & v_{d} & 0 \cr -v_{d} & 0 &
  \theta v \cr 0 & 0 
  & \theta} \hspace{.5in} {\rm and} \hspace{.5in} \lambda_U =
\pmatrix{ 0 & v_{u} e^{i\delta} & 0 \cr
 -v_{u} & 0 &  v \cr 0 & 0 & 1}.
\end{equation}

To give masses to the charm and strange quarks, we assume the presence 
of two additional bulk fields, $\xi_{u,d}$, that transform as $2 \times
 2$ under $U(2)_{Q} \times U(2)_{u^c,d^c}$\footnote{It would be
problematic to instead introduce
doublets under $U(2)_{u^c}$ and $U(2)_{d^c}$ for this purpose, 
because the relation $\theta_c \approx
\sqrt{m_d/m_s}$ would be spoiled by the Yukawa term $Q_L {{\overline
    \phi}_Q}^L {\phi_d}_l\epsilon^{lm}{D^c}_m$.  Moreover, 
the bulk coupling ${\overline \phi}_Q{\chi_d} \phi_d$ would regenerate $\chi_d$ in the
vicinity of our wall, and would also disrupt the texture (we might
expect, for instance ${\lambda_d}_{21} \sim {\lambda_d}_{22}$).}.  
However, these flavons are
not shined from distant branes, but rather have VEVs induced in the
bulk by the interactions
\begin{equation}
\mathcal{L} \supset \phi_{L} {{\overline \chi}^L}_l \phi_{ M}
  {\overline \xi}^{lM}.
\label{eq:sniffterm}
\end{equation}
Due to its sensitivity to the flavon masses and
to the brane geometry, the size of the sniffed $\xi$ is essentially
a free parameter.  Note, however, that the orientation and phase
of $\xi$ is determined entirely by the orientation and phase of
$\phi$ and $\chi$, so that we have
\begin{equation}
\xi_{u,d} = \pmatrix{0 & 0 \cr 0 & v'_{u,d}},
\end{equation}
where $v'_u$ and $v'_d$ are both real.  

Including the leading order coupling of all flavons to the quarks
yields the textures
\begin{equation}
\lambda_D = \pmatrix{ 0 & v_{d} & 0 \cr -v_{d} & v'_d &
  \theta v \cr 0 & 0 
  & \theta} \hspace{.5in} {\rm and} \hspace{.5in} \lambda_U =
\pmatrix{ 0 & v_{u} e^{i\delta} & 0 \cr
  -v_{u} & v'_u &  v \cr 0 & 0 & 1},
\label{eq:finaltex}
\end{equation}
and leads to the approximate relations
\begin{equation}
\theta_c \sim \sqrt{m_d \over m_s } \sim {v_{d} \over v'_{d}},
\hskip 0.5 in 
|{V_{ub} \over V_{cb}}| \sim \sqrt{m_u \over m_c } 
\sim {v_{u} \over v'_{u}},
\label{eq:rel1}
\end{equation}
\begin{equation}
 {m_s \over m_b} \sim {v'_{d} \over \theta}, \hskip 0.5in
{m_c \over m_t} \sim v'_{u}, \hskip 0.5in V_{cb} \sim v.
\label{eq:rel2}
\end{equation}
We get reasonable values for all observables by taking $v'_d \sim 3
\times 10^{-4}$, $v'_u \sim 3 \times 10^{-3}$, $v_d \sim 6 \times
10^{-5}$, $v_u \sim 2 \times 10^{-4}$, $v \sim 1/30$, and $\theta \sim 
1/60$.  Note that the CKM matrix is of the
form
\begin{equation}
V_{CKM} = \pmatrix{ R^u_{12} & \cr & 1} \pmatrix{ e^{i \delta} & \cr & R_{23}} 
\pmatrix{ R^d_{12} & \cr & 1}, 
\label{eq:CKM}
\end{equation}
so that the unitarity triangle relations for this model will simply be 
those of standard $U(2)$.  

How safe is this model?  In the mass basis, we expect to have the
operator of equation (\ref{eq:badop}) generated with coefficient $\sim 
(\lambda_s \theta_c)^2$, so that the bound from $\epsilon_K$ is reduced
to $\Lambda>$ 7 TeV.  The issue of the neutron EDM is more subtle.
Despite the fact that CP is broken spontaneously, one might expect
this model to have an EDM problem because the mass and 
EDM matrices are produced by several flavons, rather than by a single
multiplet as in $U(3)^5$. However, an attractive feature of this
model is that, in the mass basis, the phase $\delta$ appears only in the
CKM matrix and {\em not} in the leading order EDM matrices.  This is
clear from (\ref{eq:finaltex}):  rotating $U_1 \rightarrow e^{-i
  \alpha} U_1$ makes the  mass and EDM matrices completely real, because
the 1-2 entry of the up quark EDM matrix started out with precisely
the same phase as ${\lambda_U}_{12}$.  Higher order contributions to
these matrices disrupt this cancellation.  We find that the order
of magnitude of the contribution to the neutron EDM is determined by the coupling
\begin{equation}
Q_L {{{\overline \chi}_u}^L}_l \epsilon^{lm} {{\overline
    \chi_u}^M}_m {\xi_d}_{nM} \epsilon^{np} {D^c}_p.
\end{equation}
Setting $p=L=1$ gives 1-1 entries in both the down quark EDM and mass
matrices of roughly  $v_u^2 v'_d e^{i\delta}$ in the flavor basis.
In the mass basis, the 1-1 element of EDM matrix has approximate size
$v_d^2/v'_d$, so the phase of that element will be roughly ${v_u^2 
{v'_d}^2 \over v_d^2} \sim 10^{-6}-10^{-5}$, suggesting that the bound on 
$\Lambda$ is reduced to well below 1 TeV.   

\section{Predictive Theories}
We find the picture for generating flavor of sections 4 and 5, based on the maximal
$U(3)^5$ flavor group and involving only three identical 
flavor-breaking branes with triplet sources pointing in random 
directions, to be both elegant and plausible. 
Nevertheless, it does not provide a 
{\it predictive} theory of flavor. Furthermore, there is no explanation 
of the $b/t$ hierarchy.
There are two points that need to be 
addressed in order to build more predictive theories based on $U(3)^5$:
\vskip 0.15in
$\bullet$The brane geometry must be more constrained.
\vskip 0.08in
$\bullet$The directions in $U(3)^5$ space shone by the triplets must be more 
constrained.
\vskip 0.15in
In this section, we will present some examples of more predictive theories
along these lines. In order to deal with the second point, we will assume that
the dynamics on the branes is such that the triplet sources have
identical strengths and can only shine in
three orthogonal directions, which we take to be $(0,0,1),(0,1,0)$, and 
$(1,0,0)$. 
If we continue to work with just three parallel source branes, as in
the previous sections, we would be stuck with $V_{CKM} = 1$. 
Therefore we consider
other brane configurations.  In particular, we imagine that the
triplets $T_u$, $T_d$, and $T_Q$ are all localized on different
defects, which we label as $U^c$, $D^c$, and $Q$ branes.  We take all the 
$Q$ branes to be parallel and equally spaced, and similarly for the
$U^c$ and $D^c$ branes.  However, the $Q$ branes intersect at right
angles with both the $U^c$ and $D^c$ branes, and at the junctions
$\chi_u$ and $\chi_d$ have sources.  The Yukawa matrices 
are thus shined from the points of intersection on a grid of
flavor breaking branes.     

\subsection{Simple grid models}
\begin{figure}
  \centerline{ \psfig{file=
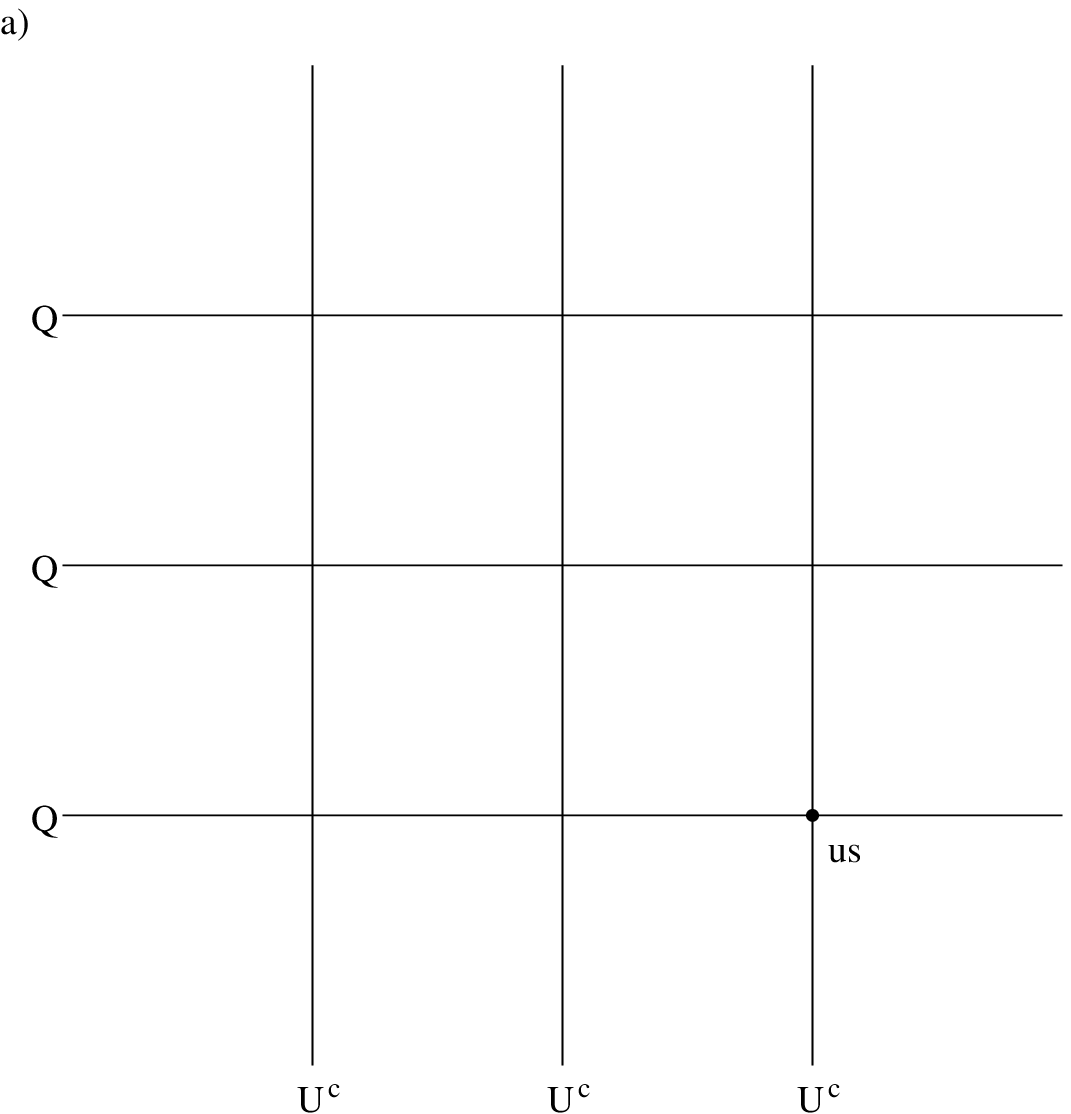,width=0.4\textwidth,angle=0}\
    \hspace{.5in}\psfig{file=
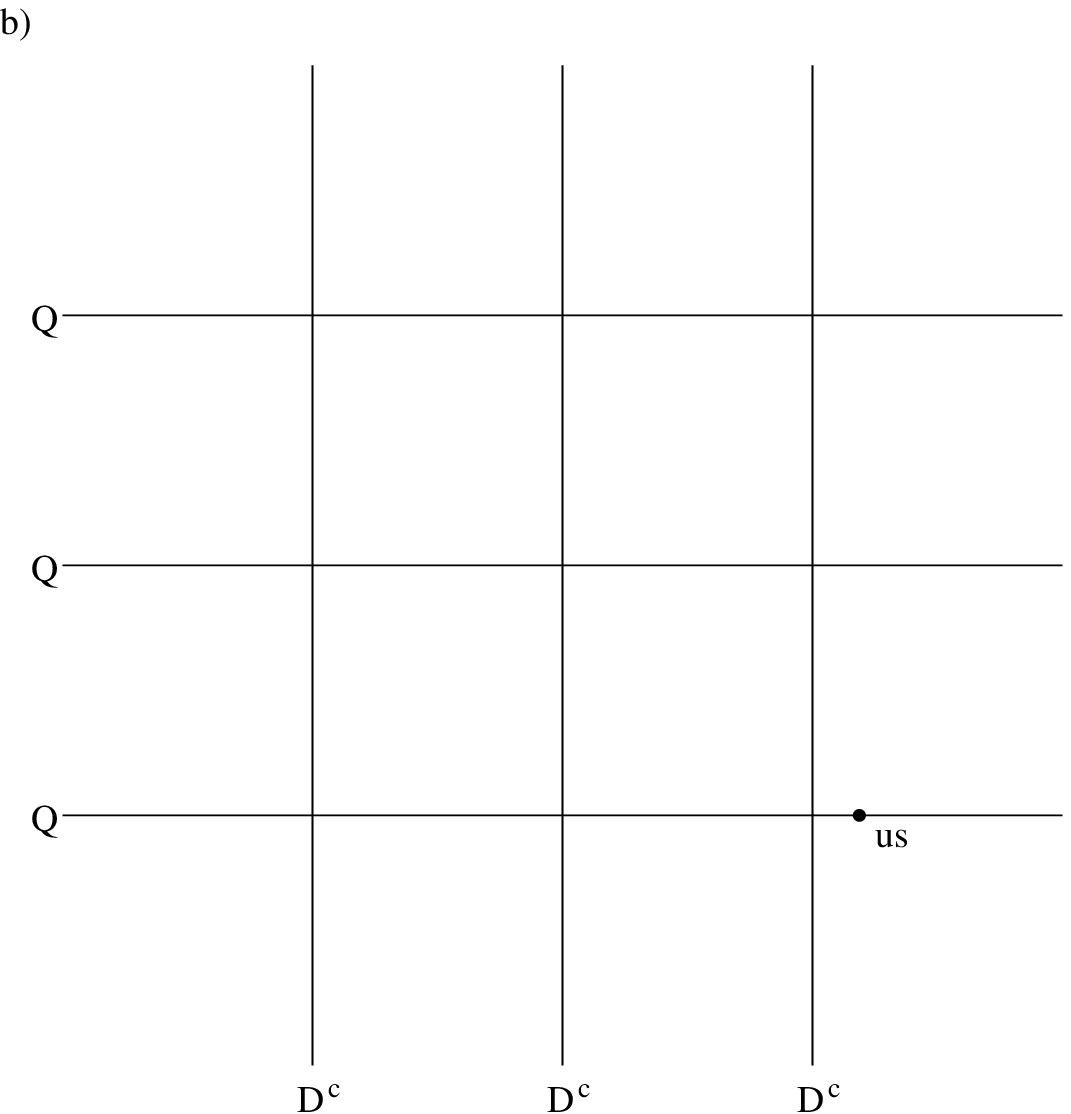,width=0.4\textwidth,angle=0} }
 \caption{The brane configuration for the first grid model.  We
   project onto a plane parallel to the 
   $D^c$ branes for (a), and parallel to the $U^c$ branes for (b).}
\label{fig:gridone}
\end{figure}

We will describe three simple and predictive
grid models.  In the first, the triplet sources are localized on
three sets of three parallel five-branes of infinite extent.
Labeling the extra dimensions by the numbers
$1,2,...,n$, we take the $Q$, $U^c$, and $D^c$ branes to fill extra 
dimensions 1 and 2, 2 and 3, and 1 and 3, respectively. At the four-dimensional
intersection of a $Q$ brane with a $U^c$ ($D^c$) brane, the $T_Q$ and
$T_u$ ($T_d$) triplets shine the bulk flavon $\chi_u$ ($\chi_d$).  We could
further imagaine the existence of an additional bulk field
$\chi_{u,d}$ that transforms as $({\overline 3,\overline 3})$ under
$U(3)_{D^c} \times U(3)_{U^c}$.  This flavon would not induce dangerous
operators on our brane, and would simply make the picture
more symmetric.  We take the masses of the three flavons, as
well as the spacings between the $Q$, $U^c$, and $D^c$ branes, to be
identical.  Until we have a theory
that determines the inter-brane separations, we can't justify the regularity
of the grid, however the symmetry of the system ensures that the configuration
is at least a local extremum of the potential.
In an attempt to understand why
$m_t \gg m_b$, we imagine that our three-brane is located at 
one of the $Q$ - $U^c$ intersections, 
but is not in contact with a $D^c$ brane.  Below 
we will find it necessary to make the offset, the shortest distance from our    
three-brane to the nearest $Q$ - $D^c$ intersection, 
much smaller than the brane spacing.  The  
configuration is represented in Figs. \ref{fig:gridone}a
and \ref{fig:gridone}b.
In Fig. \ref{fig:gridone}a  we project
onto the 1-3 plane passing through our three-brane, so that the
$Q$ - $U^c$ intersections appear as points.  In Fig. \ref{fig:gridone}b
we do the same for the 2-3 plane passing through our brane, so that the    
$Q$ - $D^c$ intersections appear as points.  We have
placed our universe near a  corner of the grid, where it is easiest to
attain hierarchical quark masses for all three generations.

We stress that the starting point for this theory is a remarkably
symmetrical configuration of source branes. In the absence of our 3
brane, and of spontaneous breakings, the configuration is completely
symmetrical with respect to interchanging any pair of the extra
dimensions 1,2 and 3. The labels $Q, U^c, D^c$ are just labels of
identical sets of branes. The lack of symmetry only occurs by virtue
of the position of our own 3 brane and the gauging on it. Flavor
symmetry is built into the large scale structure of the bulk, and is
explicitly broken only at a point defect.

Already we can see that the resulting mass matrices will have an interesting
structure. For instance, keeping only the exponential dependence of the 
$\chi$ propagators, the up mass matrix has the form
\begin{equation}
\lambda^U \sim \pmatrix{\epsilon^{2 \sqrt{2}} & \epsilon^{\sqrt{5}} & 
\epsilon^2 \cr \epsilon^{\sqrt{5}} & \epsilon^{\sqrt{2}} & \epsilon \cr
\epsilon^2 & \epsilon & 1}
\label{eq:rt2}
\end{equation}
where $\epsilon \sim$ exp $-(m_\chi S)$ is the 
suppression factor with $S$ the 
interbrane spacing. This sort of pattern is not expected in 4d theories of 
flavor, where usually only integer powers of a small parameters appear. 

We now proceed to a quantitative analysis. 
As mentioned above, we take the magnitudes of each source VEV to be
precisely the same, and allow the triplets to point in three
orthogonal directions.  We
can choose a basis where the triplets on the $Q$, $D^c$,and
$U^c$ branes nearest us all point in the 3 direction; for conciseness
we label these branes as $Q3$, $U^c3$, and $D^c3$.  The $D^c3$ - $Q3$
intersection shines the Yukawa coupling
\begin{equation}
\lambda^D_{33} = \alpha_{\chi} \, \Delta(m_\chi; y^D_{33}),
\end{equation}
where $\alpha_{\chi}$ absorbs the source strengths and their couplings
to the bulk flavons (we take these to be the same for $\chi_u$ and
$\chi_d$), and $y^D_{33}$ is the distance from our brane to the
$D^c3$-$Q^c3$ intersection.  Note that if there are $k$ extra
dimensions, the propagator is given by equation (\ref{eq:chivev})
with $n=k-1$, because the shining is from the 4D intersection of two
five-branes. Meanwhile, from the $U^c3$-$Q3$ intersection we get the
Yukawa coupling
\begin{equation}
\lambda^U_{33} =\alpha_{\chi} \, \Delta(m_\chi; y_{cutoff})+ \alpha_{T},
\end{equation}
where the second term comes from the direct coupling of the triplet
sources to standard model fields, and the
propagator has been cutoff at a distance $y_{cutoff} \sim 1/\Lambda$.
Of course, with $\alpha_{T}$, $\alpha_{\chi}$, $m_\chi$, and the offset
$y^D_{33}$, we have more than enough freedom to fit the top and bottom
quark masses; the hope is that these four free parameters plus the
brane spacing $S$ can be simultaneously chosen to give reasonable CKM mixing angles and mass
ratios for the other quarks as well.  An additional hope is that the
parameter values for a successful fit not be too far from unity.  In this
case, the smallness of $m_b/m_t$ is not put in by hand by 
simply choosing $\alpha_\chi \ll \alpha_T$, but is instead a 
consequence of our location at a $U^c$-$Q$ intersecion, away from $D^c$ 
branes. Note that even in the limit that we are very near a $D^{c}-Q$ 
interesection, $m_{b}/m_{t}$ is still suppressed by a factor if 
$\Gamma({n-2 \over 2}) / 4 \pi^{{n/2}}$.

The model we have described is quite constrained.  For a
given configuration of triplet VEVs, the quantities $y^D_{33}/S$ and
$S m_\chi$ specify all ratios of Yukawa matrix elements except those
involving $\lambda^U_{33}$; by further fixing $\alpha_{\chi}
m_{\chi}^{k-3}$, where $k$ is the number of extra dimensions,
we determine the magnitudes of all Yukawa matrix elements except $\lambda^U_{33}$,
which is given only once we choose $\alpha_T$.  Thus there are four
free paramaters to predict six masses and three mixing
angles\footnote{CP violation is discussed below.}.  
The predictions turn out to be wrong.  A qualitative reason for
this can be understood by considering only the two nearest $Q$,
$U^c$, and $D^c$ branes.  We must be able to choose the source VEVs on 
these branes as
($Q3$, $Q2$), ($U^c3$, $U^c2$), and ($D^c3$, $D^c2$) - if there were
a repetition in any of the VEV directions, then three sets of three
branes would not be sufficient to give masses to all of the 
quarks\footnote{We could have ($U^c3$, $U^c3$), and a massless up
quark, but this makes the problem described below only more severe.}.
The size of $m_s/m_b$ is approximately the
ratio of the contributions to $\chi_d$ from  $Q2$ - $D^c2$ and  $Q3$ -
$D^c3$ shining.  The distance from our brane to $Q2$ - $D^c2$
is longer than that to $Q3$ -
$D^c3$ by at least $S$, regardless of $y^D_{33}/S$; if we
make $S m_\chi$ larger than roughly 2 or 3, then $m_s/m_b$ automatically
comes out too small.  
Meanwhile the dominant contribution to $V_{cb}$ comes from the ratio
of the contributions to $\chi_d$ from $Q2$ - $D^c3$ and $Q3$
- $D^c3$ shining.  For the moderate values of of $S m_\chi$ needed
for $m_s/m_b$, getting
$V_{cb} \ll 1$ requires the offset $y^D_{33}$ to be
substantially smaller than the spacing $S$ (by roughly a factor of 3
or more).  
With $y^D_{33}/S$ constrained in this way, the ratio of the charm mass, which
arises dominantly from $Q2$ - $U^c2$ shining, to the
bottom mass, comes out too small\footnote{In the case of four extra
dimensions, for example, if we require
$.036<V_{cb}<.042$ and $1/24<m_s/m_b<1/75$, then we obtain
$m_c/m_b<1/23$, with the value of $m_c/m_b$ optimized when 
$y^D_{33}/S \sim .08$ and $S m_\chi \sim .7$.}.  Of course, this
problem can be avoided if we introduce an additional free parameter, for
instance by letting $\chi_u$ and $\chi_d$ have different masses,
by allowing the spacing between the $U^c$ and $D^c$ branes to be different,
or by giving $\chi_u$ and $\chi_d$ different couplings to the triplet
sources.  However, even if adjustments like these are made to accomodate
$m_c/m_b$, we find that it is not possible to simultaneously obtain 
accurate predictions for all other mass ratios and mixing angles.  

A simple modification of the brane grid just described is to eliminate
the $U^c$ branes and place the $T_u$ sources on the same branes as the
$T_Q$'s.  To make the picture more symmetric, we could imagine that
on the $D^c$ branes we have additional triplet sources $T_{Q'}$ that
transform under yet another $U(3)$, under which all standard model
fields are singlets.  In the original grid model, our brane needed to be located at an 
intersection of different branes to get the additional
contribution to $\lambda^U_{33}$ from the direct coupling of triplet
sources to standard model fields; here, the direct coupling is automatic
provided only that we reside on a $Q/U^c$ brane. 
Two other important
differences distinguish this grid from the previous one:  first, there
are now only three independent sources that shine $\chi_u$; second,
the classical profiles for $\chi_u$ and $\chi_d$ shining are no longer
identical - if the source branes are co-dimension $l$ objects, then 
the $\chi_u$ profile is determined using equation (\ref{eq:chivev}) with $n=l$, while
for the $\chi_d$ profile one should use $n=l+1$ (as the
intersections of $D^c$ and $Q$ branes have co-dimension $l+1$).  One fortunate effect 
of the latter difference is to increase $m_c/m_b$ from what 
the previous grid gave, for given choices of $y^D_{33}/S$, $S$, and
$m_\chi$.  

\begin{figure}
  \centerline{ \psfig{file=
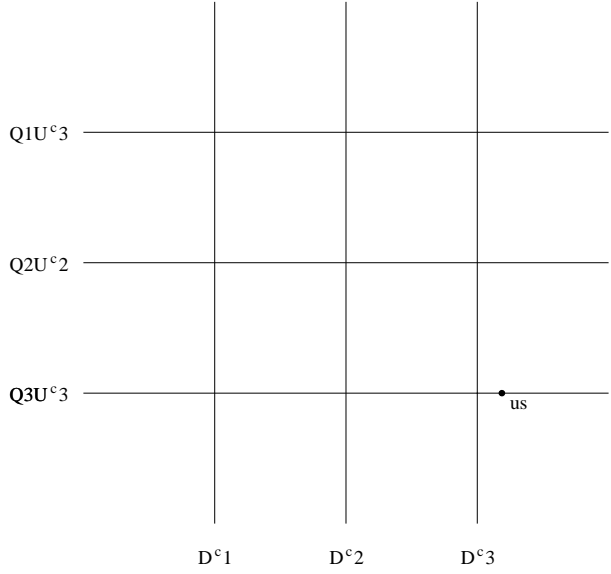,width=0.5\textwidth,angle=0}} 
 \caption{The brane configuration for the second grid model.  The
   numbers indicate the directions of the triplet source VEVs.}
\label{fig:gridthree}
\end{figure}

To determine how well this grid can fit quark masses and mixings, we need 
to specify a VEV configuration.  For all six quarks to acquire mass, 
there must not be repetitions of VEV orientations (for
example, we need one each of $D^c3$, $D^c2$, and $D^c1$).
Unfortunately, giving a 3-2-1 pattern to all three sets of triplet VEVs
leads automatically to an up quark mass that is too large compared to 
$m_d$.  In light of this we choose {\em not} to make the VEVs on the
most distant $Q/U^c$
brane ($Q1, U^c1$), but instead choose them to be ($Q1, U^c3$), as shown 
in Fig. \ref{fig:gridthree}
(choosing ($Q1, U^c2$) leads to too large a contribution to $\theta_c$ 
coming from the up sector). With this VEV configuration, the up quark
Yukawa matrix has the texture
\begin{equation}
\lambda^{U} \sim  \pmatrix{0 & 0 & \epsilon' \cr 0 & \epsilon & 0 \cr 0
  & 0 & 1}. 
\end{equation}
In particular, we have $m_u=0$, which is allowed at second
order in chiral perturbation theory\cite{kaplan}.  

Interestingly, this grid
model is slightly less constrained than the previous one:
because the $\chi_u$ and $\chi_d$ sources have different
dimensionality, the ratio $m_c/m_b$ depends on $m_\chi$ and $S$
independently, so that there are five free parameters ($m_\chi$, $S$,
$y^D_{33}$, $\alpha_{\chi}$, $\alpha_T$).  With these chosen
to be (.43, 3.3, .43, .1, $\sim$ 1)\footnote{The precise value of
  $\alpha_T$ is fixed by fitting the top mass.}, and with source branes of
co-dimension two, we obtain a mixing matrix with
elements of magnitude
\begin{equation}
V_{CKM}= \pmatrix{.975 & .223 & .0040 \cr .223 & .974 & .037 \cr .0045 
  & .037 & .999},
\end{equation}
and find masses
\begin{eqnarray}
m_d=1.9 \, {\rm MeV} \hspace{1in} m_u= 0 \hspace{.5in}\\
m_s=70 \,{\rm MeV} \hspace{1in} m_c= 1.4\,{\rm  GeV}\\
m_b=4.2 \,{\rm GeV} \hspace{1in} m_t= 174\, {\rm GeV},
\end{eqnarray}
where we have included RGE running.  
The mass ratios and the magnitudes of the CKM matrix elements 
are consistent with those inferred from data,
except that $m_s/m_d = 37$ is too high by $\sim 50 \%$, and
$|V_{ub}/V_{cb}|=.11$ is too large by $\sim 10 \%$.  Note that we are
partially successful in understanding the smallness of $m_b/m_t$:  the dimensionless parameter 
required to fit this mass ratio, $\alpha_\chi$, is $\sim 1/10$ rather
than $\sim 1/60$.  The most serious problem with the model as
presented so far is that there is no CP violation.  This is
easily remedied: if we allow the triplet VEVs to be complex,
then we are left with an irremovable phase in the CKM matrix, provided
the phases of the $T_u$ VEVs on the ($Q3,U^c3$) and ($Q1,U^c3$) branes are different.

\begin{figure}
  \centerline{ \psfig{file=
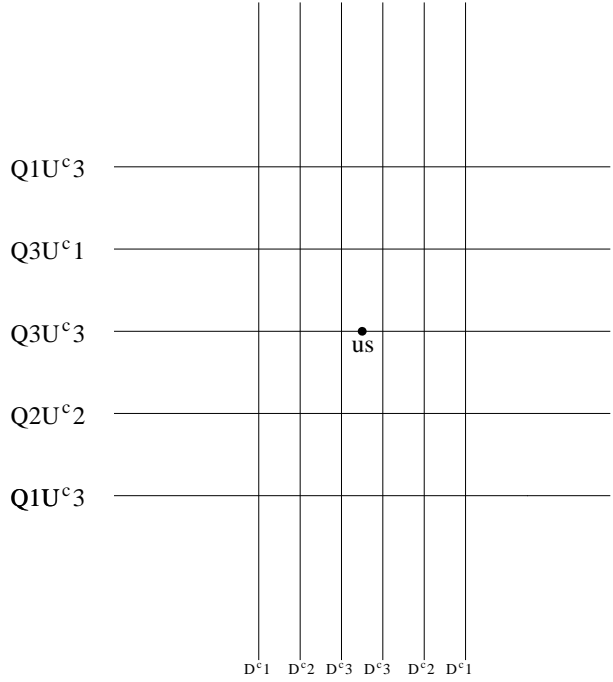,width=0.5\textwidth,angle=0}}
 \caption{The configuration of branes and source VEVs in the third
   grid model.}
\label{fig:gridfour}
\end{figure}

We find it encouraging that the simple, regular grid shown in Fig. 
\ref{fig:gridthree} 
can describe quark masses and mixings so well.  Perhaps the
strangest, least desirable feature of this grid is our peculiar
location relative to it.  This motivates the grid shown
in Fig. \ref{fig:gridfour},
in which our brane is located at the precise center. 
Given that we are in the middle, located
on a $Q/U^c$ brane and in between two $D^c$ branes, this construction
features the minimum number of branes required to give
masses to all three down-type quarks. 
We have eliminated the free parameter $y^D_{33}$; in its place we
allow the $D^c$ brane spacing to differ from the spacing of the
$Q/U^c$ branes.  With the orientation of source VEVs shown in
Fig. \ref{fig:gridfour}, the up quark Yukawa matrix is
\begin{equation}
\lambda^{U} \sim  \pmatrix{0 & 0 & \epsilon' \cr 0 & \epsilon & 0 \cr \epsilon
  & 0 & 1}, 
\end{equation} 
so that now the up quark obtains a small mass proportional to
$\epsilon \epsilon'$.  If we choose our free parameters
($m_\chi$, $S_{D^c}$, $S_{Q/U^c}$, $\alpha_{\chi}$, $\alpha_T$) to be
(.23, 2.75, 8, .2, $\sim$ 1), then we obtain 
 \begin{equation}
V_{CKM}= \pmatrix{.976 & .219 & .0057 \cr .219 & .975 & .039 \cr .0030 
  & .039 & .999},
\end{equation}
and 
\begin{eqnarray}
m_d=4.3 \,{\rm MeV} \hspace{1in} m_u=1.3 \,{\rm MeV}\\
m_s=130 \,{\rm MeV} \hspace{1in} m_c=1.3 \,{\rm GeV}\\
m_b=4.3 \,{\rm GeV} \hspace{1in} m_t=174 \,{\rm GeV}.
\end{eqnarray}
All masses and mixings agree with experiment at the $50 \%$ level:
 $|V_{td}|$ and $m_d/m_s$ are both too small by $\sim 25 \%$, while
$|V_{ub}|/|V_{cb}|$ is too large by $\sim 50 \%$.  Again, we find
it intriguing that the symmetric grid of Fig. \ref{fig:gridfour}, 
with our 3 brane
at its center, can account rather well for the pattern
of quark masses and mixings.

\section{Conclusions}
The gauge hierarchy problem has motivated several directions for
constructing theories beyond the standard model. 
Each of these has presented certain 
challenges and opportunities for making progress on the flavor problem. 
Constructing realistic theories of fermion masses in technicolor theories 
without fundamental scalars proved to be very \, difficult - especially 
incorporating the heavy top quark. In the simplest supersymmetric 
theories, the Yukawa couplings of the standard model are simply copied as 
superpotential interactions. As in the standard model there is an 
economical description which provides no understanding of the origin of 
flavor. The ideas for understanding the origin of small dimensionless 
Yukawa couplings are the same as for theories without supersymmetry: 
perturbative loops or the Froggatt Nielsen mechanism using hierarchies of 
mass scales. While there are new twists on these old ideas -- 
superpartners can be in the loop and the dynamics of many strongly 
interacting supersymmetric theories are understood -- at the end of the 
day one is tempted to say that supersymmetry did not allow much progress 
in understanding flavor. In fact,  for supersymmetric theories the 
question has been how to avoid taking a step in the wrong direction: there 
are severe constraints from flavor-changing and CP violating processes on 
the form of the soft supersymmetry breaking interactions involving squarks 
and sleptons. While the answer motivated some flavor groups, it may be 
that these constraints are telling us more about how supersymmetry is 
broken than about how flavor is broken.  

In contrast, if we live on a three brane at some location in the bulk, 
with the fundamental scale from our viewpoint of order a TeV, then the 
constraints on theories of flavor are radically altered, and a whole new 
world of flavor models is opened up. At first sight it again appears that 
we are heading in the wrong direction: how could disastrous flavor 
changing effects be avoided from operators generated at
such a low scale, from 
familons and from light flavon exchange? We have argued that all three 
objections are immediately removed by having a discrete non-Abelian flavor 
group spontaneously broken on source branes in the bulk. The fundamental 
scale of flavor breaking on these source branes is order unity, but the 
breaking effects on our 3-brane are small because the source branes are 
distant from us. 
The origin of the flavor parameters is now a convolution of two effects: 
the geometrical configuration of the source branes in the bulk and our 
location relative to them, and the random relative orientations of the 
flavor breaking vevs on the various source branes. Phenomenology places some 
constraints on these effects, and energetics suggest that the brane 
configuration will be highly symmetrical. We find that the convolution of
these two effects has sufficient complexity to lead to the collection of
mystifying flavor numbers nature has given us, while still originating
from a
very simple and elegant symmetrical structure. An interesting aspect of
this  picture is that flavor symmetry is a crucial feature of the
extra dimensions 
and is important in determining the brane configurations in the bulk. On 
the other hand, our gauge interactions are restricted to our 3-brane, and 
are unimportant from the viewpoint of the bulk. 

It is important to stress that these theories really are new, and cannot 
be mimicked by 4 dimensional theories.
For example, the relative size of entries in the Yukawa matrices are
governed by distances to sources from which flavor breaking is
shone. These relative distances involve factors, such as $\sqrt{2}$ and
$\sqrt{5}$ as shown in eqn. [\ref{eq:rt2}], which are characteristic 
of the spatial geometry. Furthermore, these theories of flavor can
occur whether the extra dimensions are large, small or infinite, and
whether the background geometry is flat or curved.
In addition, there are new ideas for a
qualitative explanation of features of the fermion mass spectrum. In
grid theories a fermion mass hierarchy is {\it inevitable} -- our 
3-brane must be located closer to some source branes than to others. The
uniquely heavy top quark is explained by having our 3-brane located on
source branes which break flavor symmetry in the up sector. The origin
of $m_t / m_b$ may be a brane configuration such as the one shown in
Figure 8, where our 3-brane lies equidistant between two source branes
for breaking flavor in the down sector.

As well as a new structure for flavor, and new ideas for qualitative
features of the fermion mass spectrum, theories in extra dimension
offer a completely new mechanism for obtaining precise flavor
parameter predictions. It is striking that there have been only a few
theoretical ideas which lead to relations amongst the flavor
observables, such as $\theta_c \approx \sqrt{m_d/m_s}$ and $m_b
\approx 3 m_\tau$. Texture zeros, symmetry properties of the Yukawa
matrices, and grand unified relations between up, down and lepton
sectors have been the most important tools. Extra dimensions offer a
completely new possibility: the Yukawa matrices on our 3-brane at
$y_0$ are given by $\chi(y_0) = \sum_i  \chi_i \Delta(y_0 - y_i)$
where $y_i$ is the location of source $i$ which shines the flavor
breaking $\chi_i$ via the propagator $ \Delta(y_0 - y_i)$. The
positions $y_i$ depend only on the lattice spacing, and there may be
few possible spontaneous choices for the orientation of $\chi_i$. 
This basic idea can be implemented in a wide range of models.

There are clearly very many source brane structures to be considered, even 
concentrating on those with high symmetry, and one may question whether 
such constructions are plausible origins for the quark and lepton mass 
matrices. We find little reason to prefer the alternative picture of multiple 
Frogatt-Nielsen fields and flavons with masses enormously high compared to 
the TeV scale.  Rather than debate the relative merits, it seems worth 
exploring this new class of theories in which there is a 
spatial geometry of flavor.

\section*{Acknowledgements}
We would like to thank S. Dimopoulos for useful discussions.
This work was supported in part by the U.S.\
Department of Energy under contract DE--AC03--76SF00098 and by the
National Science Foundation under grant PHY--95--14797.

\end{document}